\documentclass[12pt,a4paper]{article}

\usepackage[margin=2cm]{geometry}

\usepackage{graphicx}
\usepackage[inkscapelatex=false]{svg}
\graphicspath{{./figures/}}

\usepackage{titling}
\usepackage{xcolor}
\usepackage{amsmath}
\usepackage{authblk}
\usepackage{soul}
\usepackage{makecell}
\usepackage{subcaption}
\usepackage{caption} 

\usepackage[style=nature]{biblatex}
\AtEveryBibitem{
  \clearfield{doi}
  \clearfield{url}
  \clearfield{note}
  \clearfield{urldate}
  \clearfield{ISSN}
  \clearfield{issn}
  \clearfield{urlyear}
  \clearfield{urlmonth}
}

\author[1,2]{D. A. Poteryayev}
\author[3]{P. Novelli}
\author[1,3]{A. Coriolano}
\author[4]{R. Dettori}
\author[5]{V. Tozzini}
\author[2]{F. Beltram}
\author[3]{M. Pontil}
\author[1]{A. Rossi\thanks{Corresponding author: antonio.rossi@iit.it}}
\author[1]{S. Forti\thanks{Corresponding author: stiven.forti@iit.it}}
\author[1]{C. Coletti\thanks{Corresponding author: camilla.coletti@iit.it}}

\affil[1]{Center for Nanotechnology Innovation @NEST, Italian Institute of Technology, Pisa 56127, Italy}
\affil[2]{NEST Laboratory, Scuola Normale Superiore, Italy, Pisa 56127, Italy}
\affil[3]{Computational Statistics and Machine Learning, Italian Institute of Technology, Genova 16163, Italy}
\affil[4]{Department of Physics, University of Cagliari, Cagliari 09042, Italy}
\affil[5]{Nanoscience Institute, National Research Council (CNR-NANO), Pisa 56127, Italy}
\date{}

\begin{document}

\begin{refsection}
\title{SpectraFormer: an Attention-Based Raman Unmixing Tool for Accessing the Graphene Buffer-Layer Signature on SiC}



\maketitle

\section*{Abstract}

Raman spectroscopy is a key tool for graphene characterization, yet its application to graphene grown on silicon carbide (SiC) is strongly limited by the intense and variable second-order Raman response of the substrate. 
This limitation is critical for buffer layer graphene, a semiconducting interfacial phase, whose vibrational signatures are overlapped with the SiC background and challenging to be reliably accessed using conventional reference-based subtraction, due to strong spatial and experimental variability of the substrate signal.
Here we present SpectraFormer, a transformer-based deep learning model that reconstructs the SiC Raman substrate contribution directly from post-growth partially masked spectroscopic data without relying on explicit reference measurements. 
By learning global correlations across the entire Raman shift range, the model captures the statistical structure of the SiC background and enables accurate reconstruction of its contribution in mixed spectra.
Subtraction of the reconstructed substrate signal reveals weak vibrational features associated with ZLG that are inaccessible through conventional analysis methods.
The extracted spectra are validated by \textit{ab initio} vibrational calculations, allowing assignment of the resolved features to specific modes and confirming their physical consistency.
By leveraging a state-of-the-art attention-based deep learning architecture, this approach establishes a robust, reference-free framework for Raman analysis of graphene on SiC and provides a foundation, compatible with real-time data acquisition, to its integration into automated, closed-loop AI-assisted growth optimization.

\section{Introduction}

Graphene is an exceptional two-dimensional (2D) material that has become a platform for groundbreaking fundamental physics studies on Dirac fermions, quantum Hall physics, and many-body interactions \cite{castro_neto_electronic_2009}. Its flakes can be obtained through mechanical exfoliation of graphite crystals, and these samples have historically set the benchmark for the highest electronic quality.

Despite the remarkable scientific achievements enabled by exfoliated flakes, the “Scotch-tape” method inherently limits graphene research to micrometer scale samples and low throughput fabrication. Consequently, one of the central challenges for the scientific community has been to scale up graphene synthesis while maintaining the material quality required to probe fundamental physics and enable its technological integration. Among the various approaches explored, chemical vapor deposition (CVD) on copper has emerged as a widely used technique capable of producing large-area graphene with electronic properties approaching those of exfoliated flakes \cite{gebeyehu_decoupled_2024, pezzini_high-quality_2020, li_large-area_2011, banszerus_ultrahigh-mobility_2015}. However, monolayer graphene (MLG) is not a semiconductor, as it lacks an energy band-gap, which leads to a low On/Off state current ratio. Such property limits graphene use in field-effect transistors and other logic devices. Overcoming this limitation remains one of the key challenges for translating graphene’s properties into electronic applications.

In some case scenarios, it is possible to induce a band-gap in graphene based materials using various approaches: giant in-plane and/or out-of-plane strain \cite{gui_reply_2009, wong_strain_2012, pereira_tight-binding_2009, ni_uniaxial_2009}, electromagnetic field \cite{mak_observation_2009, xia_graphene_2010, avetisyan_stacking_2010}, close interaction with nearby materials \cite{xu_interfacial_2018}, or any combination of them in order to break the lattice pseudospin symmetry. Graphene grown by thermal decomposition of hexagonal silicon carbide (nH-SiC) uniquely provides an \textit{intrinsically wafer-scale} platform for high-quality graphene \cite{emtsev_towards_2009}. On the Si-terminated SiC$(0001)$ surface, this method proceeds through the sublimation of Si atoms, leaving behind a carbon-rich reconstruction with a $(6\sqrt{3}\times6\sqrt{3})\,R30^\circ$ periodicity \cite{van_bommel_leed_1975}, where one third of the carbon atoms are covalently bonded to the substrate. This layer, known as the buffer layer or zero-layer graphene (ZLG), features a graphene-like lattice but lacks the electronic structure of graphene due to its bonding to the substrate \cite{emtsev_interaction_2008, goler_revealing_2013, cavallucci_intrinsic_2018}. It exhibits semiconducting properties (0.4-0.6 eV band gap \cite{n_nair_band_2017, zhao_ultrahigh-mobility_2024}) due to its partial sp$^3$ hybridization and strong interaction with the substrate.

While enormous attention has been devoted to the resulting graphene layers, which have also become an established platform for quantum electrical metrology \cite{tzalenchuk_towards_2010, janssen_quantum_2013, wan_contrasting_2023}, ZLG itself has recently gained interest as a promising material for semiconductor technology, and it is emerging as an important platform for stabilizing confined two dimensional materials at the SiC interface, including two dimensional gold \cite{forti_semiconductor_2020} and other two dimensional metals \cite{briggs_atomically_2020}.

Despite its potential, reproducibly growing high-quality ZLG on wafer scale remains difficult. The process depends sensitively on various growth protocol's parameters, which are often interdependent. Controlling these parameters with precision is essential for tuning the electronic properties and achieving uniformity across the substrate. Traditional growth optimization relies on trial-and-error approach, which is slow and inefficient for such a multidimensional parameter space, along with its high cost of probing. Artificial intelligence (AI) provides a data-driven route to accelerate this process by autonomously exploring and optimizing the growth parameters.

Raman spectroscopy remains one of the most powerful and widely used tools for identifying graphene \cite{ferrari_raman_2013} owing to its speed, robustness, and ease of use. Furthermore, we recently proved Raman spectroscopy to be highly compatible with closed-loop AI-assisted growth optimization \cite{sabattini_towards_2025}. However, Raman analysis of the ZLG presents a significant challenge: its spectral features are entirely obscured by the dominant SiC signal, composed by bands originating from ion-implantation damage as well as overtone of the folded acoustic and optical modes \cite{nakashima_raman_1997}. Traditional approaches rely on subtracting a reference SiC spectrum, but this process is highly sensitive to experimental conditions such as focus, exposure, and sampling position. Even minor deviations lead to unreliable subtraction, making it impractical for real-time or automated growth pipelines, unlike free standing graphene, where there is still 2D peak that can be used for material characterization. These factors limit the ability to use Raman data for real-time feedback during growth optimization.

\begin{figure}[!htb]
    \centering
    \includegraphics[width=1.0\linewidth]{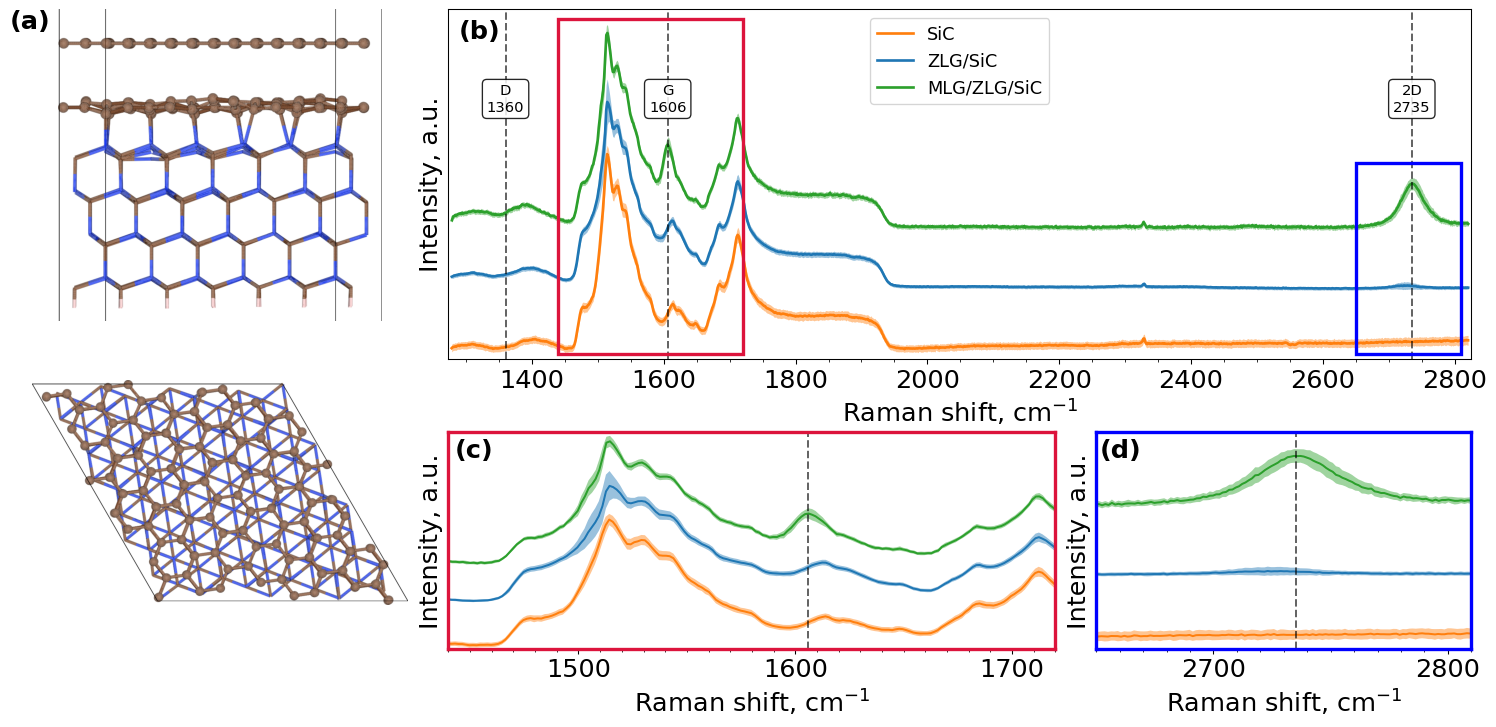}
    \caption{(a) Ball-and-stick atomic structure model of investigated materials (SiC substrate, ZLG and  MLG); (b) Raman spectroscopy of 3 different samples (solid line - mean across the dataset, shaded area - 1 standard deviation (STD) area around mean value): bare SiC substrate, SiC substrate with ZLG, and SiC substrate with both ZLG and free standing MLG. From insets (c,d) it is easy to see the appearance of surface material's signal. Measurement conditions are all same for all represented spectra: 5\% laser power, 5 seconds acquisition time, 1 accumulation per spectrum.}
    \label{fig:ball-and-stick-comparison}
\end{figure}%

To address this limitation, we propose leveraging on transformer architectures~\cite{turner2023introduction}, a class of deep-learning models originally developed for natural language processing \cite{vaswani_attention_2023}. Transformers are AI models operating on sequences, which excel at identifying non-local correlations in data, making them particularly well suited for spectroscopic tasks. In this work, we introduce SpectraFormer, a transformer-based AI model, to unmix the ZLG Raman signal from the SiC background without explicit reference subtraction. The model treats the Raman signal as an ordered sequence of intensity values, and it outputs the reconstructed SiC spectrum directly from the given part of mixed signal.

We further validate our model using \textit{ab-initio} calculations on a high-fidelity supercell model that captures vibrational features previously overlooked \cite{radtke_vibrational_2025}. These theoretical insights align with the spectral contributions extracted by the transformer, confirming both the physical reliability of our method and the presence of buffer-layer signatures in experimental spectra.

The approach presented here offers a new route for real-time, robust Raman analysis during graphene growth on SiC, bypassing the limitations of reference-based subtraction and enabling integration into closed-loop AI-driven synthesis workflows. More broadly, our results highlight the \textit{transformative} potential of attention-based models for spectral unmixing in materials characterization, paving the way for accelerated discovery and scalable manufacturing of next-generation 2D materials.

\section{Materials and methods}

ZLG is grown through thermal decomposition, starting from a 6H-SiC substrate. Commercial SiC wafers are first thoroughly cleaned using acetone and IPA in an ultrasonic bath, then subjected to oxygen plasma treatment, followed by Piranha and HF baths to eliminate any organic residues from the surface. Subsequently, the SiC(0001) wafers are exposed to high-temperature hydrogen gas inside a furnace to remove polishing scratches, a procedure known as hydrogen etching. Once this step is completed, the ZLG growth process begins. All growth procedures are conducted using an AIXTRON Black Magic cold-wall reactor in an Ar environment.

Grown ZLG on SiC samples were further characterized using X-ray photoemission spectroscopy (XPS), confirming the presence of ZLG related features (for details, see Supplementary). 

All Raman spectroscopy data were acquired using Renishaw inVia micro-Raman spectrometer. Since the region of interest for ZLG features is located in range (1000 -- 3000) cm$^{-1}$, datasets were acquired on both 6H and 4H-SiC(0001) polytypes in that range centered at different positions (centers were chosen in range from 1800 cm$^{-1}$ to 2300 cm$^{-1}$ with 100 cm$^{-1}$ step) with span of 1500-1600 cm$^{-1}$ for each spectrum (1015 data points per spectrum), using 100x objective with 0.85 NA, 532 nm excitation laser and 1800 lines/mm grating. For balancing training data and enriching model knowledge, different measurement parameters were used. In particular, different combinations of: acquisition time, laser power, and number of accumulations per spectrum (Fig.~\ref{fig:distribution} in Supplementary).

Model training was conducted on 4 NVIDIA Tesla V100 16Gb GPUs using Franklin HPC infrastructure of IIT. Model training took 1:16:39 of computation time (1.28 hours). To train SpectraFormer AI model, conventional Adam optimizer was used \cite{kingma_adam_2017}. Model was trained on arithmetically preprocessed Raman spectroscopy data (preprocess pipeline is described in Supplementary). Preprocessing was a necessary step to address several problems, such as presence of outliers and numerical artifacts appearance during calculations. 

We model the ZLG/SiC interface using the reduced commensurate supercell often denoted as the $\sqrt{31}\times\sqrt{31}\,R8.95^\circ$ reconstruction of the Si-terminated SiC surface, which corresponds to a $7\times7$ graphene-like overlayer rotated by $R21.787^\circ$. In this work, we adopt the same structure considered in \cite{Cavallucci2018, Cavallucci2018Carbon}: the \emph{Sh} stacking, characterized by the presence of at least one "hollow" site (a surface Si atom located beneath the center of a buffer hexagon). The initial configuration is generated by placing a flat hexagonal carbon layer above the Si-terminated surface at a separation slightly larger than the expected Si-C bond length and then fully relaxing atomic positions (see Computational details below). The reduced supercell enables calculations at significantly lower computational cost than the larger $6\sqrt{3}$-based model, while retaining the key local structural motifs and registry patterns relevant to the ZLG. The substrate is modeled as a four-layer slab of cubic SiC, with the bottom surface passivated by hydrogen atoms to remove spurious dangling-bond states.

All calculations are performed within Density Functional Theory (DFT), with the same setup adopted in previous works \cite{Cavallucci2018, Cavallucci2016} using ultrasoft RRKJ pseudopotentials \cite{Rappe1990PRB} and the PBE exchange-correlation functional \cite{Perdew1996PRL}. Dispersion interactions are included through the semi-empirical Grimme D2 correction (PBE-D2) \cite{Grimme2006JCC}. The plane-wave kinetic-energy cutoff is set to 30~Ry, with a charge-density cutoff of 300~Ry, and self-consistency is converged to $10^{-8}$.

Structural relaxations are carried out with a quasi-Newton BFGS optimizer \cite{Billeter2003CMS}, using standard convergence thresholds of $10^{-3}$~a.u.\ on forces and $10^{-4}$~a.u.\ on total energy. Geometry optimizations and SCF calculations to extract the forces from the atomic displacements are performed at the $\Gamma$ point. A Gaussian smearing of 0.01~Ry is used throughout.

The in-plane lattice vectors are fixed to those of a relaxed SiC slab, while the out-of-plane supercell length is set to 31.8~\AA\ to avoid interactions between periodic replicas. Calculations are performed with \textsc{Quantum ESPRESSO} \cite{Giannozzi2009JPCM} (version 7.2.0).

Interatomic force constants and vibrational modes have been obtained using the \textsc{ALAMODE} \cite{alamode} code within the frozen-phonon approach. Since the ZLG hosts in-gap and near-gap electronic states that are spatially localized on specific atomic motifs (crest regions, benzene-like patches, etc.), we limited atomic displacements to these regions while keeping the remainder of the simulation cell fixed to reduce computational cost. See Supplementary Information for more details.

\section{Results}
To clarify the physical origin of the spectral unmixing problem addressed in this work, we first consider the structural and spectroscopic evolution of graphene grown on SiC during thermal decomposition.
Fig.\ref{fig:ball-and-stick-comparison}a presents a schematic ball-and-stick representation of the investigated system:
a 4H-SiC(0001) substrate gets deprived by its Si atoms by thermal annealing. In particular, when three SiC bilayers are decomposed, the carbon atoms in excess re-arrange on the surface and form the ZLG. By further annealing at higher temperatures, deeper layers of the substrate lose their Si atoms, and a new ZLG forms, transforming the former one into monolayer graphene. The latter one is now only weakly interacting with the substrate, represented by the ZLG/SiC(0001) and it therefore develops all the electronic properties typical of graphene. In the ZLG instead one carbon atom out of three is covalently bound to the substrate. Hence, the interatomic hopping between carbon atoms is stronlgy suppressed and so are the $\pi$-bands, which indeed are not developed in the ZLG. The vibrational properties of MLG and ZLG are substantially different as well. 
Fig.\ref{fig:ball-and-stick-comparison}b shows representative Raman spectra acquired from three distinct sample configurations: bare SiC, ZLG/SiC, and MLG/ZLG/SiC. In all cases, the Raman response is dominated by the intense second-order features of the SiC substrate, which extend over a broad spectral range and mask weaker carbon layers contributions. As the ZLG forms, additional spectral components emerge; however, their direct identification remains challenging due to the overwhelming SiC signal and the partial overlap of vibrational modes.
Fig.\ref{fig:ball-and-stick-comparison}c and Fig.\ref{fig:ball-and-stick-comparison}d focus on spectral regions where carbon layers contributions are most apparent. In the presence of a graphene monolayer, the characteristic graphene 2D peak remains clearly observable, enabling straightforward identification of MLG despite the dominant substrate signal overlapping with expected D and G peaks positions. In contrast, for the ZLG, graphene-derived features are strongly suppressed. A weak residual 2D-like signal can nevertheless be detected, which is attributed to small graphene enclosures or localized regions of incomplete surface conversion that can form during ZLG growth. These localized graphene inclusions do not correspond to a continuous graphene overlayer, but they contribute a faint graphene-like signature to the overall Raman response.

As mentioned above, while the presence of a MLG can still be identified through its characteristic 2D peak, the vibrational signatures of the ZLG are almost entirely masked by the dominant and broadband Raman response of the SiC substrate. Importantly, this masking effect is not fixed but varies between measurements as a result of slight changes in experimental conditions, such as focus, alignment, and acquisition parameters, as well as local crystalline defects, to which the SiC band is associated in that region \cite{wang_raman_2019}. As a consequence, reference-based background subtraction or local fitting strategies become unreliable and difficult to apply in a systematic or automated manner. Addressing this challenge requires an analysis strategy that does not rely on a predefined reference spectrum and does not assume that spectral information can be extracted from local regions alone. Instead, the relevant information is distributed across the entire Raman spectrum and across many measurements, and must be inferred from their collective statistical structure. 

Transformer-based models provide a suitable framework for this task by treating each Raman spectrum as an ordered sequence of intensity values and learning how different parts of the spectrum are statistically related to one another. In practice, this is achieved by first mapping the chosen raw spectral intensities into a learned representation space, referred to as spectral embeddings, which allows the model to encode complex spectral patterns beyond simple peak positions or amplitudes. The model then applies self-attention mechanism, which quantify how variations in intensity at one Raman shift are correlated with variations at all other Raman shifts within the same spectrum. Through multiple stacked layers of this mechanism, the model progressively builds a global representation of the substrate Raman response, capturing both strong features and subtle correlations that are not apparent from local inspection. By exploiting these global and non-local correlations, the transformer is able to infer the most probable contribution of the SiC substrate even in spectral regions that are intentionally hidden during training. This capability fundamentally distinguishes the approach from local interpolation, polynomial fitting, or convolution-based methods, which primarily rely on neighboring spectral points. As a result, transformer-based models are particularly effective at disentangling weak, overlapping interfacial contributions embedded within a strong and structured background, enabling robust spectral unmixing under variable experimental conditions \cite{ghosh_hyperspectral_2022}.

\begin{figure}[!htb]
    \centering
    \includegraphics[width=0.95\linewidth]{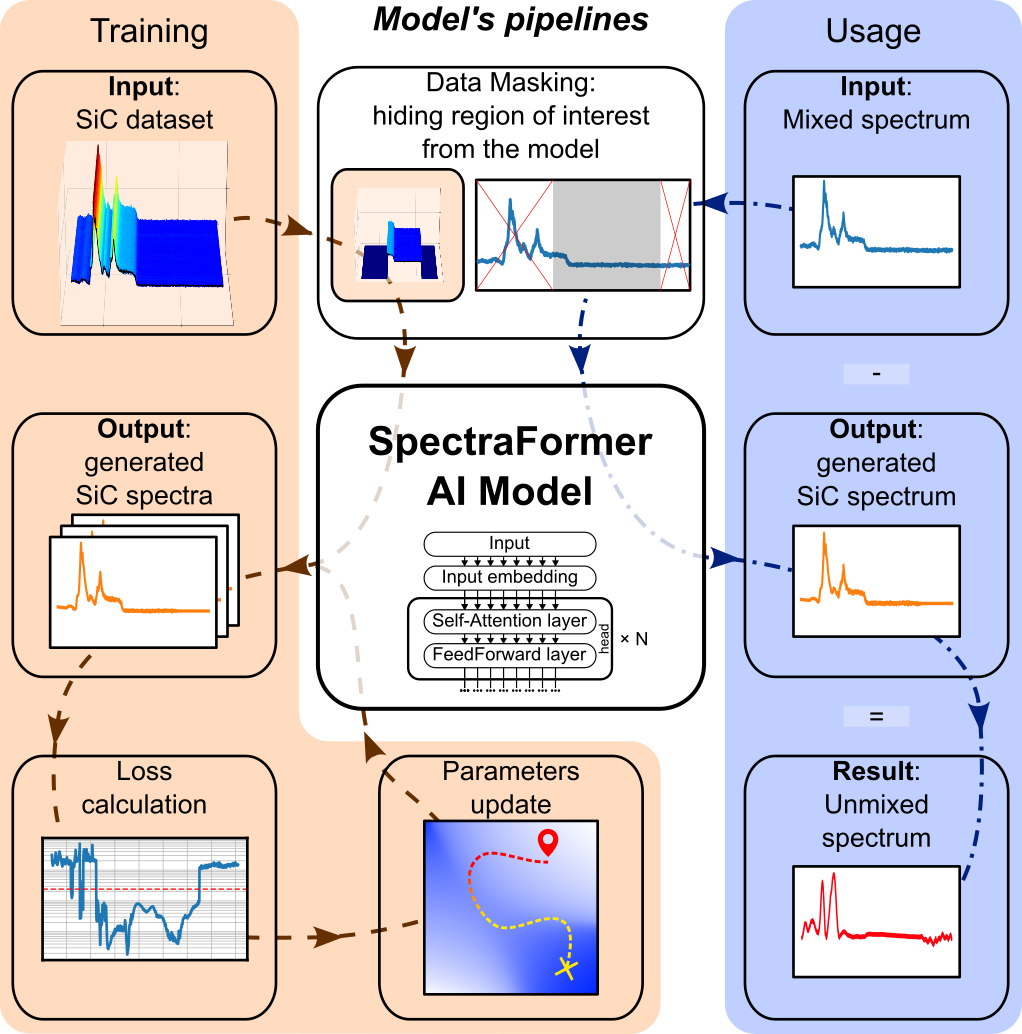}
    \caption{Model's pipelines for training and usage cases. For both approaches the model generates a bare SiC substrate signal reconstruction based on masked input, but it differs in the input type (for training - SiC spectra, for usage - mixed spectra) and whether model parameters being updated or not.}
    \label{fig:pipeline}
\end{figure}%

Based on this representation, the training and inference pipelines of the SpectraFormer model are schematically illustrated in  Fig.\ref{fig:pipeline}. The two pipelines share the same forward-pass through the transformer architecture but differ in whether model parameters are updated.
During training, batches of experimentally acquired SiC Raman spectra are first partially masked along the Raman shift axis. The masked regions correspond to spectral intervals that are intentionally hidden from the model and define the region of interest for the unmixing task. This masking strategy prevents the model from trivially reproducing the input signal and instead forces it to infer the missing spectral content from the remaining visible portions of the spectrum.
The masked spectra are then passed through the SpectraFormer model, which generates a reconstructed SiC spectrum over the full Raman shift range. The predicted output is compared to the original, unmasked spectrum to compute the loss function. The loss is evaluated across both the Raman shift and sample index dimensions, ensuring that the model learns not only individual spectral features but also their statistical variability across the full dataset. For clarity, Fig.\ref{fig:pipeline} illustrates the loss calculation along the Raman shift axis for a single spectrum, while the full loss aggregation procedure is described in the Supplementary Information.

\begin{figure}
    \centering
    \includegraphics[width=0.5\linewidth]{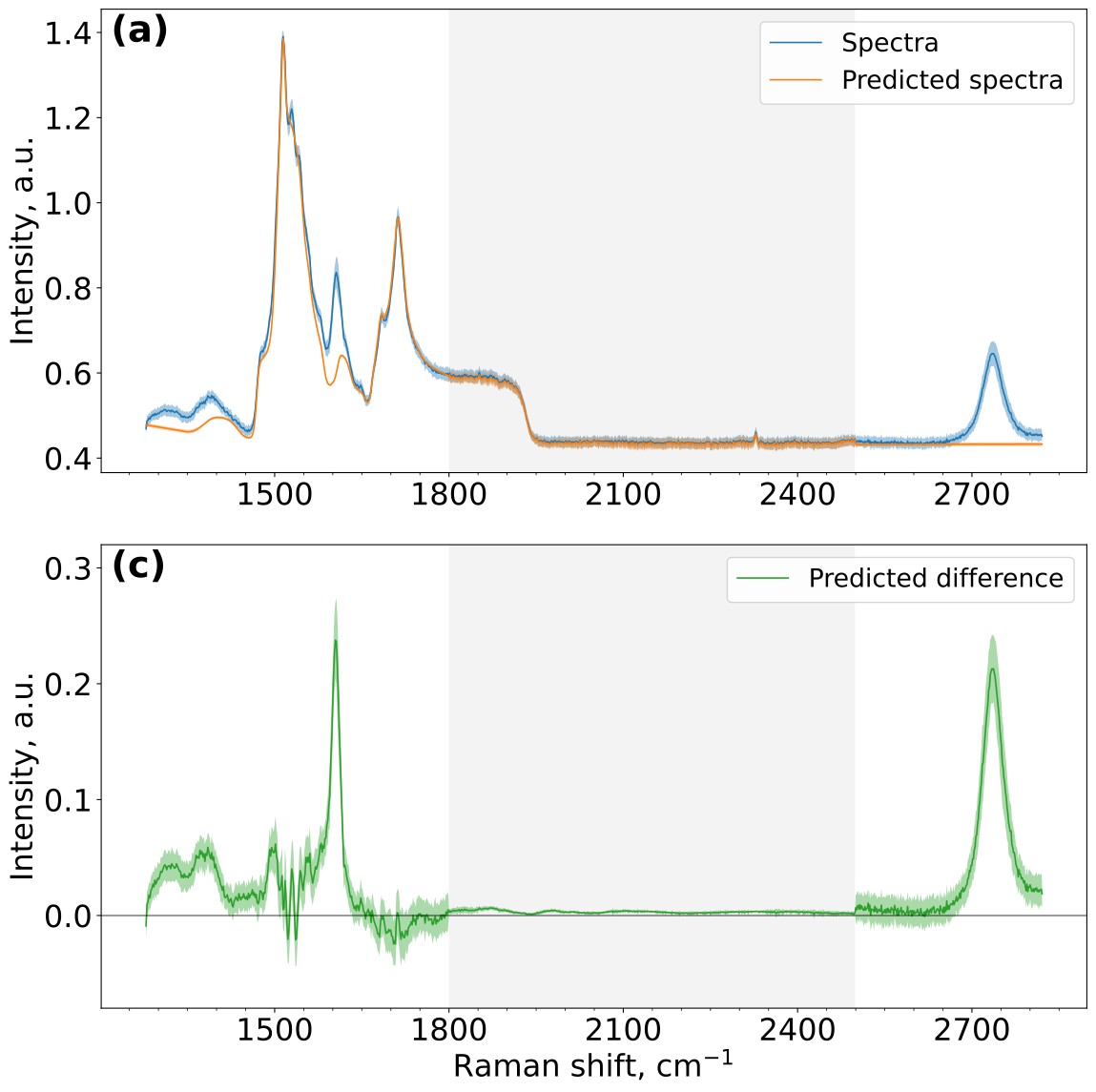}\includegraphics[width=0.5\linewidth]{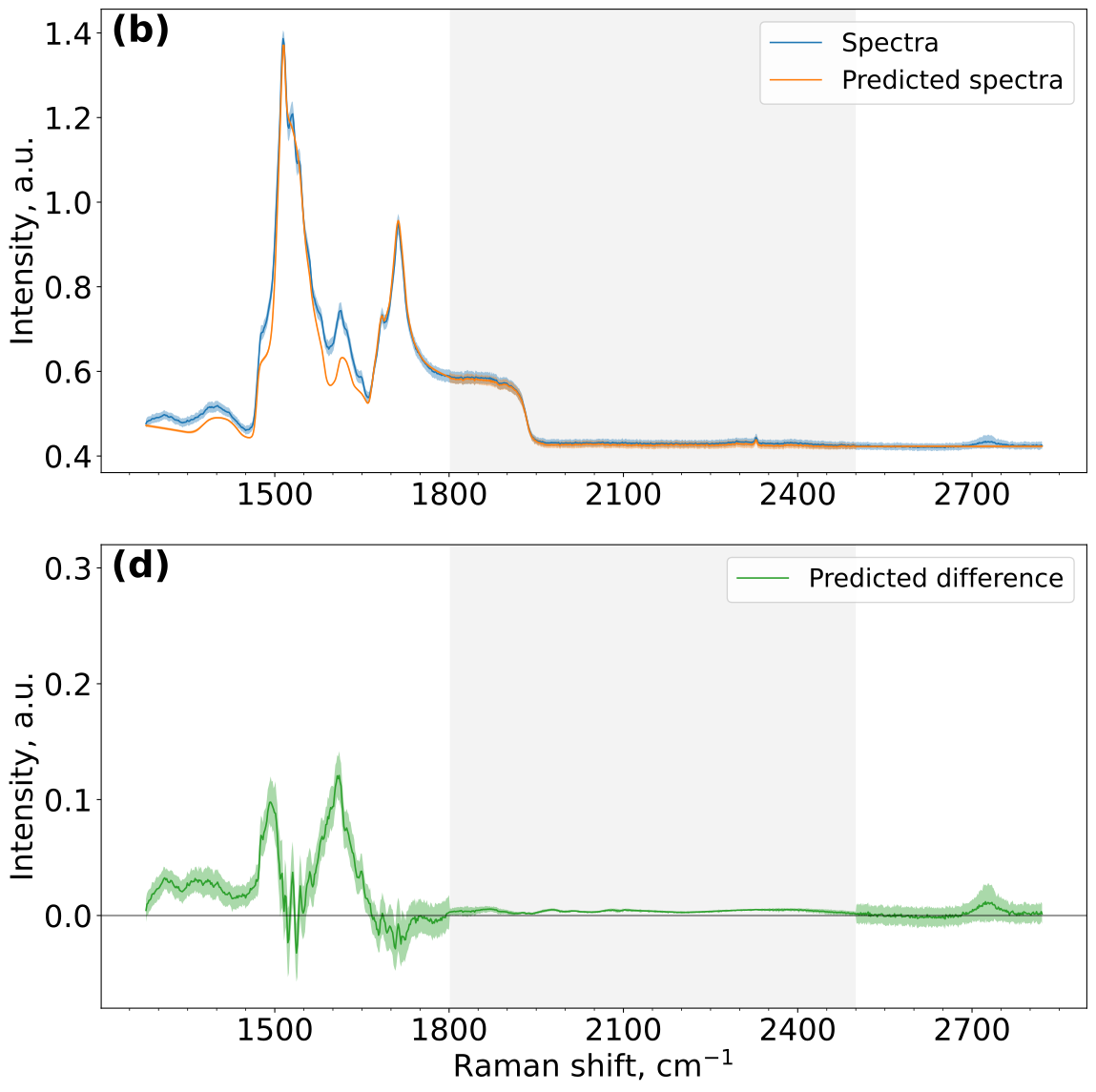}
    \caption{Model output after the training with different inputs (solid line - mean across the dataset, shaded area - 1 STD area around mean value): (a, c) MLG/ZLG/SiC Raman spectrum is given and (b, d) ZLG/SiC Raman spectrum is given, allowing to reveal targeted features by subtraction of generated SiC spectrum; middle gray shaded region is the region of data available to the model. Intensity values at (c, d) can be interpreted as normalized to the SiC peak at 1514 cm$^{-1}$, while at (a, b) are also shifted by +0.4 a.u. (for details of data preprocessing, see Supplementary).}
    \label{fig:model_outputs}
\end{figure}

Having established how the transformer learns and reconstructs the substrate Raman response, we now turn to its application to experimental spectra, where the trained model reveals Raman features of materials grown on SiC that differ from bulk SiC by capturing the substrate contribution with high accuracy. Fig.\ref{fig:model_outputs} shows the result of transformer-based unmixing applied to Raman spectra of graphene-related layers on SiC and demonstrates that this approach enables a controlled isolation of the SiC contribution, thereby exposing spectral features that are obscured in conventional measurements.

For both sample configurations (Fig.\ref{fig:model_outputs}a for MLG/ZLG/SiC and Fig.\ref{fig:model_outputs}b for ZLG/SiC), the reconstructed SiC spectra are consistent over the entire Raman-shift range, indicating that the model captures the substrate contribution independently of the overlying carbon layers. After subtraction of the reconstructed SiC signal, the residual spectra contain only non-SiC contributions. Once the dominant substrate background is removed, spectral features associated with carbon layers become clearly visible, most prominently in the low Raman-shift region. These features are consistent with Raman signatures previously attributed to the ZLG \cite{fromm_contribution_2013, radtke_vibrational_2025} and are revealed without manual background modeling or peak selection.

\section{Discussion}

It is crucial that transformer-based unmixing provides direct systematical experimental outcome (Fig.\ref{fig:model_outputs}), where the SiC substrate contribution is reconstructed and subtracted from mixed Raman spectra, giving access to vibrational features previously overlapped with the substrate ones and being directly unresolvable in conventional Raman spectra, dominated by the SiC background, without relying on reference subtraction or manual background modeling. 

A direct comparison of the two configurations provides internal validation of the unmixing. In the ZLG/SiC case (Fig.\ref{fig:model_outputs}d), the residual spectrum shows weak graphene 2D peak that is attributed to a minor signal originating from the initial stages of graphene growth, consistent with the absence of a fully developed MLG. In contrast, clear G and 2D contributions appear in the MLG/ZLG/SiC case (Fig.\ref{fig:model_outputs}c), reflecting the presence of the graphene monolayer. The relative intensity of 2D and G features differs from that expected from pristine undoped graphene. However, this behavior is consistent with the strong charge doping known for epitaxial graphene on SiC \cite{sabattini_towards_2025}, for which the 2D/G intensity ratio can vary over a wide range \cite{ bruna_doping_2014}. Noticeably, G and 2D peaks of MLG are recovered  only when a graphene monolayer is present, while the absence of a 2D peak in the ZLG-only case demonstrates that the model does not artificially introduce monolayer graphene-associated features.

To validate the features identified in the unmixed ZLG/SiC spectrum, we computed the vibrational modes of ZLG using DFT-based lattice-dynamics calculations. We extracted the one-phonon Raman response of the ZLG using a Raman-like proxy similar to the approach recently introduced by Radtke and Lazzeri \cite{Radtke2025}. Specifically, the ZLG vibrational atomic displacements at $\mathbf{q}=0$ are projected onto the Raman-active optical modes of graphene ($E_{2g}$, $\omega_{\rm DFT} = 1538.2~\mathrm{cm}^{-1}$), and the resulting overlap is used as a relative Raman-like weight (see SI for further details). This construction isolates the component of ZLG vibrational dynamics that is most graphene-like while naturally incorporating the effects of ZLG corrugation and partial covalent coupling to the SiC substrate. In addition to the B and L features, the calculated spectrum reveals a distinct $E_{2g}$-like contribution in the G-frequency range (Fig.~\ref{fig:theory_S3}b). An enhanced G contribution is expected when a significant fraction of the probed structure retains stronger graphene-like vibrational character, and its apparent visibility and linewidth can depend on the degree of structural organization and inhomogeneity \cite{Radtke2025}. The emergence of this G component in our analysis indicates that a non-negligible portion of ZLG vibrations preserves quasi-graphene-like character despite the presence of partial $sp^3$ bonding and out-of-plane corrugation. 

The calculated spectrum (Fig.~\ref{fig:theory_S3}b), is characterized by three distinct peaks, with peak positions listed in Table~\ref{tab:dft_raman}, along with their corresponding experimental values. B peak frequency is in excellent agreement with the experimental fitted one ($\omega^{\rm exp}_B=1492.3~\mathrm{cm}^{-1}$), while peaks L and G instead appear slightly red-shifted with respect to experiment ($\omega^{\rm exp}_L=1563.6~\mathrm{cm}^{-1}$ and $\omega^{\rm exp}_G=1607.0~\mathrm{cm}^{-1}$). Overall, the level of agreement indicates that the present Sh model provides an accurate structural representation of the ZLG/SiC interface; the small deviation is plausibly linked to the electronic-structure description, in particular to the choice of exchange-correlation functional. In this frequency range, semilocal GGAs are known to soften C-C stretching vibrations, leading to an underestimation of high optical phonons in diamond and graphite/graphene-related systems \cite{Mounet2005,Nakano2021}.

\begin{figure}
    \centering
    \includegraphics[width=0.95\linewidth]{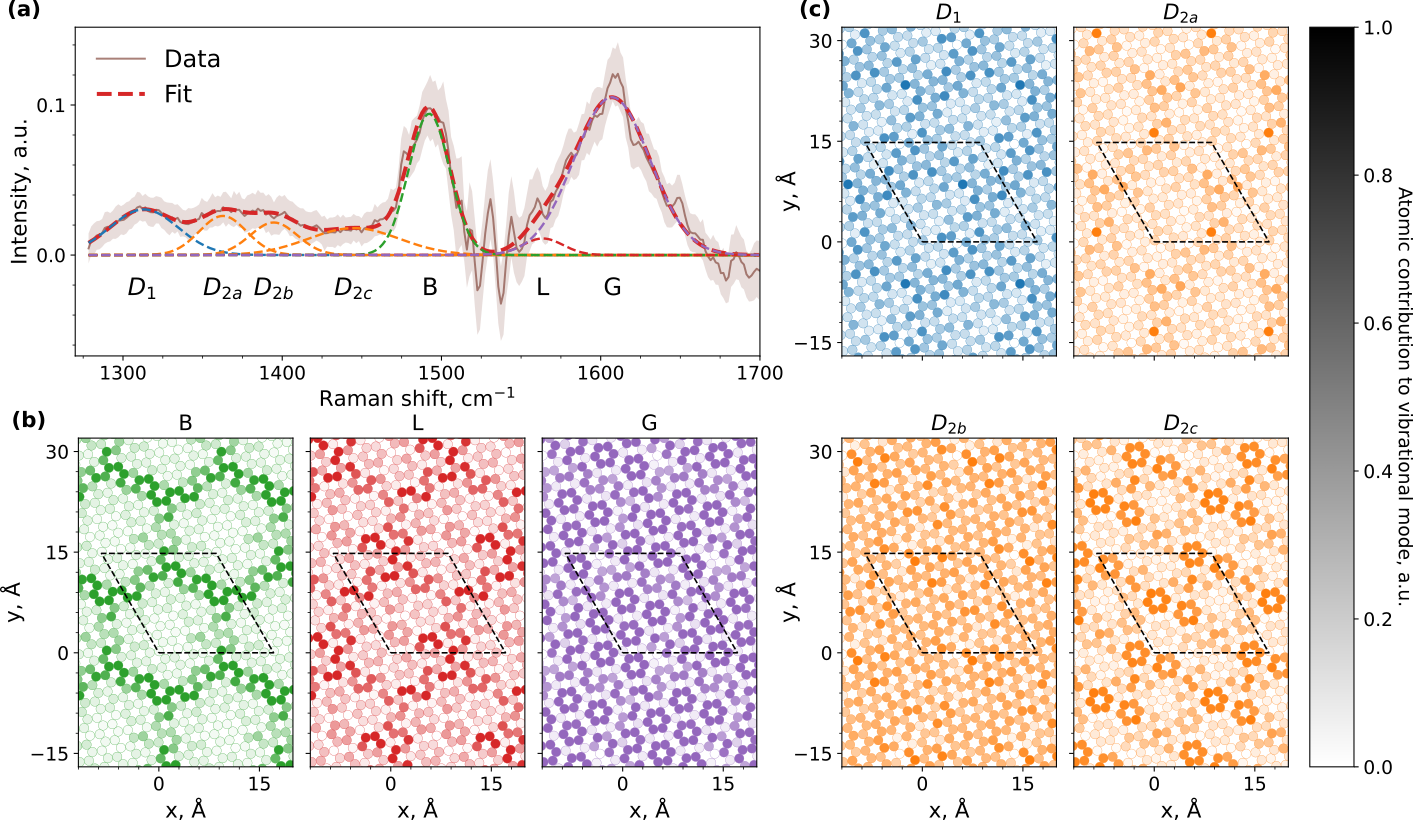}
    \caption{(a) Zoomed region of experimental data in Fig.\ref{fig:model_outputs}d with fit (solid line - mean across the dataset, shaded area - 1 STD area around mean value), where each component is taken from \textit{ab initio} calculations; (b,c) Spatial maps of atomic contribution to each of vibrational modes used (with shared gray color bar).}
    \label{fig:multipanel}
\end{figure}

Because the Raman-like intensity is obtained via the squared sum of the projection of ZLG displacement patterns onto the $E_{2g}$ modes of graphene, the spectrum can be partitioned into local atomic contributions. The integration of the atom-resolved weights over the frequency windows associated with the three peaks yields the real-space maps shown in Fig.~\ref{fig:multipanel}b, which represent the microscopic origins of the B, L, and G features. The B peak intensity follows the superperiodic crest network, pointing to a collective vibration of the elevated crest framework with mixed $sp^2/sp^3$ character from the buffer/substrate coupling, as shown also by the height map and bond distributions plots reported in Fig.~a,b. The L contribution is strongly non-uniform and dominated by sparse hotspots, consistent with a more localized vibration associated with locally strained $sp^2$ regions (e.g., distorted rings and proximity to Si-C bonded sites), in agreement with previous observations \cite{Radtke2025}. Finally, the G peak is comparatively uniform across the extended $sp^2$ network, with a reduced weight on the substrate-bonded (more $sp^3$-like) intruding atoms, reflecting a more canonical in-plane stretching character.

Unlike the B/L/G features, the D band is not expected to emerge from a purely $\Gamma$-point
Raman-like construction, since its Raman intensity is activated by symmetry breaking and
involves finite-$\mathbf{q}$ phonons through defect- and disorder-assisted processes \cite{Tuinstra1970,Venezuela2011,Ferrari2000}. Accordingly, the D band is absent in the $\mathbf{q}=0$ $E_{2g}$-projection spectrum, consistent with the discussion in Ref.~\cite{Radtke2025}. As shown in Ref.~\cite{Radtke2025}, the ZLG phonon spectrum shows an enhanced activity of the higher optical modes in the proximity of the M and K points. To interpret the experimental D-region in Fig.~\ref{fig:multipanel}a, we turn to the full vibrational density of states (vDOS), rather than the Raman-like $\Gamma$-point proxy.
For $\omega>1000~\mathrm{cm}^{-1}$ the vDOS is dominated by buffer-layer vibrations (see Fig.~\ref{fig:theory_S2}), which makes it a suitable descriptor for assigning the D-band components inathis frequency range. In the corresponding D-region of the computed vDOS (see Fig.~\ref{fig:theory_S3}a), we identify a lower-frequency component (D$_1$) and a broader, higher-frequency D$_2$ band that can be resolved into multiple subcomponents as in the experimental spectrum (D$_{2a}$-D$_{2c}$). The calculated frequencies align very well with the experimental peaks (see Tab.~\ref{tab:dft_raman}). To visualize the microscopic origin of these D-band components, we compute atom-resolved maps by weighting the squared atomic displacement amplitudes by their corresponding degeneracy and integrating over the frequency windows associated with D$_1$ and D$_{2a}$-D$_{2c}$ (see SI for further details). While these maps are intrinsically noisier than those obtained from the Raman-like proxy (because they collect contributions from all vibrational modes within the selected windows), they still provide clear qualitative trends (Fig.~\ref{fig:multipanel}c). The D$_1$ contribution is relatively diffuse, indicating the involvement of extended buffer regions shaped by substrate coupling and local corrugation (see height and Si-C bonding maps in Fig.~\ref{fig:theory_S5}a,c). By contrast, the higher-frequency components become progressively more heterogeneous and dominated by localized hotspots, consistent with a superposition of nearby vibrational modes within the reconstructed D band. The most localized patterns are associated with highly strained environments at tile boundaries and near Si-bonded sites (D$_{2a}$, D$_{2b}$), and with strongly distorted or intruded ring motifs (D$_{2c}$). Overall, Fig.~\ref{fig:multipanel}b,c support a robust physical interpretation of the experimental data: B reflects collective crest vibrations, L fingerprints localized strained $sp^2$ motifs, G tracks the extended $sp^2$ network (Fig.~\ref{fig:theory_S5}b), and the D-band captures a mixture of disorder-activated, finite-$\mathbf{q}$ vibrations whose atomistic signatures are distributed across multiple local environments rather than a single uniform motif \cite{Tuinstra1970,Venezuela2011,Maultzsch2004,Ferrari2013}.

\begin{table}[]
    \centering
    \begin{tabular}{|c|c|c|c|c|c|c|c|}
        \hline
       Vibrational mode  &  B & L & G & D$_1$ & D$_{2a}$ & D$_{2b}$ & D$_{2c}$\\
       \hline
       \makecell{Position\\(\textit{ab initio}), cm$^{-1}$}  &  1490.1 & 1542.4 & 1585.7 & 1276.6 & 1364.3 & 1390.9 & 1430.3\\
       \hline
       \makecell{Position\\(experimental), cm$^{-1}$}  &  1492.3 & 1563.6 & 1607.0 & 1313.1 & 1362.7 & 1394.7 & 1445.0\\
       \hline
    \end{tabular}
    \caption{Values of experimental and theoretical peak positions discussed for the ZLG/SiC case}
    \label{tab:dft_raman}
\end{table}

Overall, we demonstrated that transformer-based spectral unmixing and first principles modeling are complementary rather than competing approaches. SpectraFormer enables experimental access to weak interfacial vibrational signatures that are otherwise obscured by dominant substrate backgrounds, thereby making direct comparison with \textit{ab initio} predictions possible. This synergy resolves long-standing ambiguities in the Raman interpretation of the ZLG on SiC and establishes a general strategy for extracting physically meaningful information from complex, background-dominated spectra. More broadly, the same framework can be extended to other material systems where strong, structured substrates hinder spectroscopic analysis, enabling automated and reliable Raman-based assessment of interfacial layers and providing a practical foundation for closed-loop, AI-assisted optimization of growth and processing conditions.

\section{Conclusions}

In this work, we addressed the challenge of spectral unmixing in Raman spectroscopy of graphene on SiC, specifically the inability to reliably access the vibrational signatures of ZLG due to the intense, structured, and experimentally variable Raman response of the SiC substrate. 
We introduced SpectraFormer, a transformer-based deep learning architecture that is trained on partially masked spectra to reconstruct the SiC substrate contribution, allowing us to disentangle the weak vibrational signatures of the interfacial ZLG from the dominant substrate background.
Unlike traditional subtraction methods, the use of which is limited under varying experimental conditions, our approach leverages the self-attention mechanism to learn global, non-local correlations within the spectral data, enabling accurate reconstruction in mixed experimental spectra. Subtraction of the reconstructed substrate signal consistently reveals weak vibrational features associated with ZLG. 
The accuracy of this unmixing is supported by \textit{ab initio} vibrational calculations, confirming that the revealed features correspond to genuine vibrational excitation modes.
By enabling robust, automated identification of ZLG quality, SpectraFormer addresses a key challenge in the scalable fabrication of epitaxial graphene. It provides feedback for real-time, closed-loop AI-driven growth optimization. Broadly, our results establish attention-based models as a robust framework for materials characterization, capable of extracting physical insights from complex, noise-dominated spectroscopic regimes where conventional analysis is insufficient.

\section*{Declaration of Competing Interest}

Authors declare no conflict of interest.

\section*{Acknowledgments}

We acknowledge the project PNRR MUR Project PE000013 CUP J53C22003010006 Future Artificial Intelligence Research (FAIR) and PNRR MUR Project PE0000023 - National Institute of Quantum Science and Technology (NQSTI) funded by the European Union - NextGenerationEU. 
\\
We acknowledge that the research activity herein was carried out using the IIT Franklin HPC infrastructure; we gratefully acknowledge the Data Science and Computation Facility and its Support Team for their support and assistance on the IIT High Performance Computing Infrastructure.
\\
We acknowledge the CINECA award under the ISCRA initiative, for the availability of high-performance computing resources and support under project IsCd0\_SPHERE HP10CSHBY3.


\printbibliography[title=References]
\end{refsection}

\begin{refsection}
\setcounter{figure}{0}
\setcounter{table}{0}
\setcounter{section}{0} 

\renewcommand{\thefigure}{S\arabic{figure}}
\renewcommand{\thetable}{S\arabic{table}}

\title{Supplementary Information for\\ "SpectraFormer: an Attention-Based Raman Unmixing Tool for Accessing the Graphene Buffer-Layer Signature on SiC"}



\emptythanks
\maketitle

\stepcounter{section}%
\section*{Supplementary Note \thesection: XPS measurement fitting procedure}

\begin{figure}[!ht]
    \centering
    \includegraphics[width=0.85\linewidth]{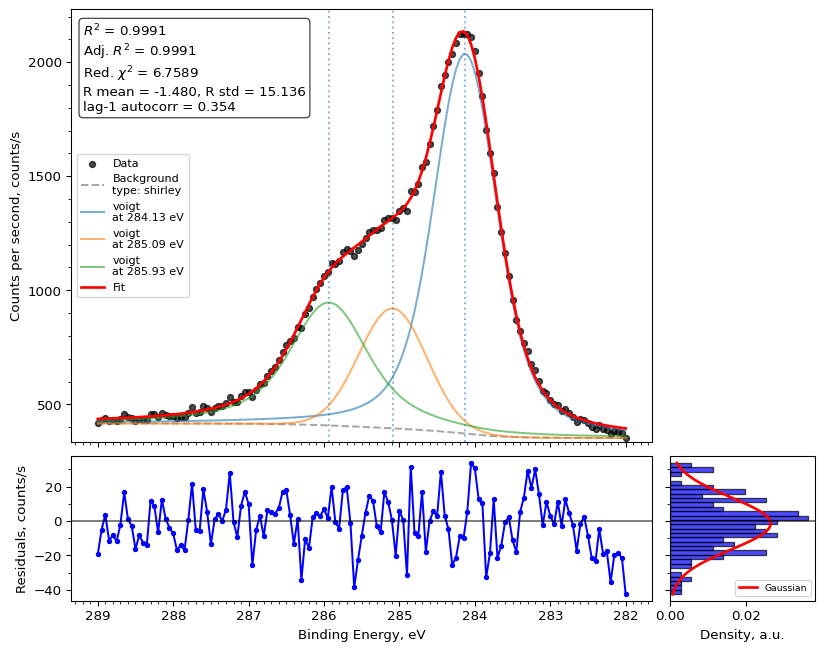}
    \caption{XPS measurement of ZLG/SiC sample.}
    \label{fig:XPS}
\end{figure}

X-ray photoelectron spectroscopy (XPS) spectra are obtained with a SPECS  XR-50 Al K$\alpha$ X-ray source. An XPS measurement performed on the ZLG sample in carbon 1s-region shows the appearance of 3 peaks (Fig.~\ref{fig:XPS}). A Shirley-type background is considered. The most prominent peak is coming from carbon signal of the bulk SiC. Two minor peaks represents an envelop of other carbon electronic configurations within a sample, including: sp$^3$- and sp$^2$-hybridized carbon atoms of ZLG itself that is exposed, graphene monolayer sp$^2$-hybridized carbon atoms (since there is a graphene contamination of the sample), and 2 additional ZLG components, to also account for the parts of ZLG that interact with MLG. Unfortunately, the resolution did not allow us to reveal all these features. Area ratio of those minor peaks is $\sim40\%$, giving a rough estimation of sp$^3$- to sp$^2$-hybridized carbon atoms ratio in ZLG.

\stepcounter{section}%
\section*{Supplementary Note \thesection: Model training}

\subsection{Statistical parameters of SiC Raman data}

\subsubsection{Data content}

\begin{figure}[!ht]
    \centering
    \includegraphics[width=0.85\linewidth]{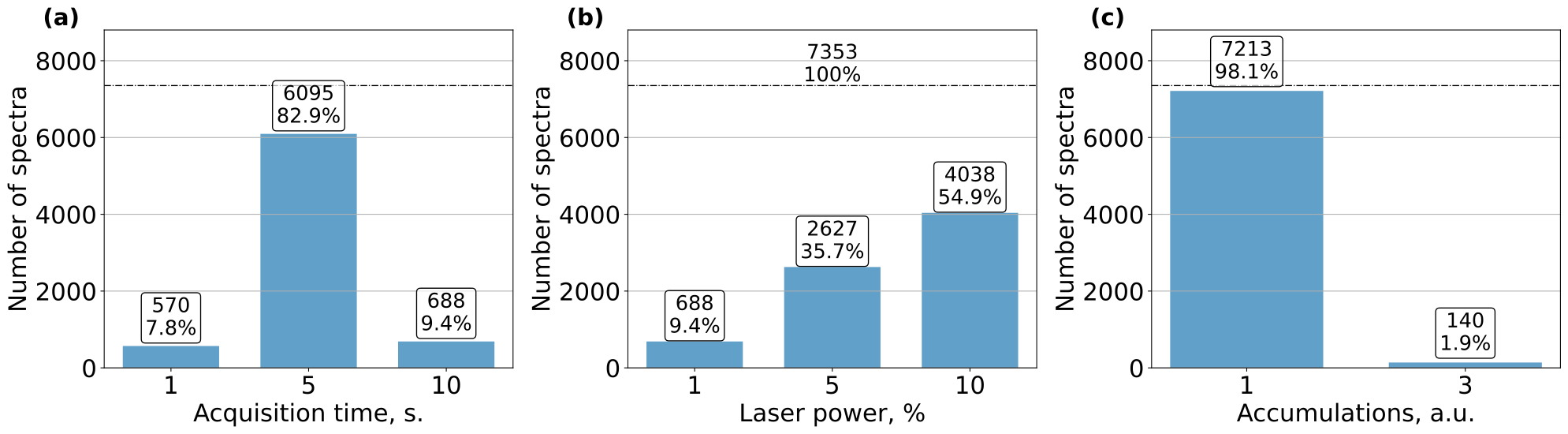}
    \caption{Statistical distribution of all spectra used for model training across different measurement parameters: (a) acquisition time, (b) laser power, (c) number of accumulations. Horizontal dashed line on each graph is the mark of total amount of spectra allowed to be seen by the model during training process.}
    \label{fig:distribution}
\end{figure}

Total amount of spectra used for the training were 7554 in total, with 80\%:20\% division for training and validation sub-datasets. But only 7353 spectra were actually provided to the model because of the outlier firewall in data preprocessing pipeline. In this work, we aimed for unmixing relatively noisy spectra, which is quite easy to obtain. Therefore, the majority of spectra were acquired with time of 5 seconds and with only 1 accumulation (Fig.~\ref{fig:distribution}). However, the noise itself also introduce uncertainty in the result of unmixing process in the region of interest directly. Because of that, a compromise was found, which of compensating the low time with higher laser power values, keeping this degree of freedom as unbiased as possible.

\subsubsection{Data preprocessing}

It is very important to have a preprocessing step for model training in ML, since data can be biased, contain outliers, and be very high and/or low in absolute value, which is hard for computations, leading to a number of computational artifacts, including variable overflow or appearance of \textbf{NaN} values. To address it, several steps in preprocessing were made:
\begin{enumerate}
    \item Spatial dimension stacking: from equipment's "X\_0", "X\_1" (and for depth maps, also "X\_2") spatial coordinates to "spectra" -- to ensure spatial independence on the prediction;
    \item Cosmic rays (outliers) removal: by applying Whitaker-Hayes algorithm based on modified Z-score \cite{whitaker_simple_2018};
    \item Spectrum normalization: the following formula is applied for each intensity value within each spectrum: \( \frac{I_i-\min I_i}{\max I_i} \);
    \item Value shifting by adding arbitrary empirical value \(I_i+0.4\) to each intensity point for each spectrum. This step is added after several numerical experiments, and aims for better loss minimum convergence, avoiding too steep curvature on the hyperplane narrow region;
    \item Outlier removal 2: a firewall with the purpose to ensure no outlier is bypassed into the model training process. Consists of several steps:
    \begin{enumerate}
        \item For each dataset (which is now 2-dimensional: "wavenumber" and "spectra") median value across "spectra" dimension is calculated. 
        \item Then, sub-norm deviations are defined as 
        
        \( \mathrm{dev}_{spectra} = \max_{spectra}\left( \operatorname{abs} \left( I_i - \operatorname{median} I_{spectra} \right) \right) \);
        \item Those spectra within a dataset that does not satisfy the condition \(\mathrm{dev}_i<0.15\) (an arbitrary empirical threshold) are dropped from the dataset.
    \end{enumerate}
    \item Wavenumber transformation: to ensure the convergence of all the calculations it is necessary to transform also the wavenumber values in the region close to the \(\left(-1;1\right)\) range. It is done by applying transformation \( \frac{\lambda_i-2000}{800}  \) with arbitrary empirical coefficients taken from region of interest values. For example, right end of spectrum at \(\sim2800\) cm$^{-1}$ will be mapped to 1, while left end of spectrum at \(\sim 1250\) cm$^{-1}$ to \(-0.9375\). For low frequency data the smallest value \(\sim0\) cm$^{-1}$ after such transformation is \(-2.5\), stating that this transformation is sufficient even for a different region of interest in low frequency range for future tasks. The only problem could be in extremely high frequency range, that usually is never probed.
\end{enumerate}

\subsection{Loss function}

In Machine Learning (ML) one is free to use any loss function \(L\) for model adjustment during training process. However, if there is some primary knowledge about data statistical behavior -- then a loss function can be defined accordingly with use of \textbf{regression analysis}, leading to convergence to the lower minimum  on model parameters hyperplane \(L(\theta)\).
    
In our case, the best-suited Loss function is the one obtained from Gamma distribution:

\begin{equation}
L=\operatorname{mean} \left[ L_i\left(I_i^{true}, I_i^{prediction}\right) \right] = \left(\frac{I_i^{true}}{I_i^{prediction}}-1\right)-\ln{\left(\frac{I_i^{true}}{I_i^{prediction}}\right)}
\label{eq:gamma_loss}
\end{equation}

\begin{figure}[!ht]
    \centering
    \includegraphics[width=0.85\linewidth]{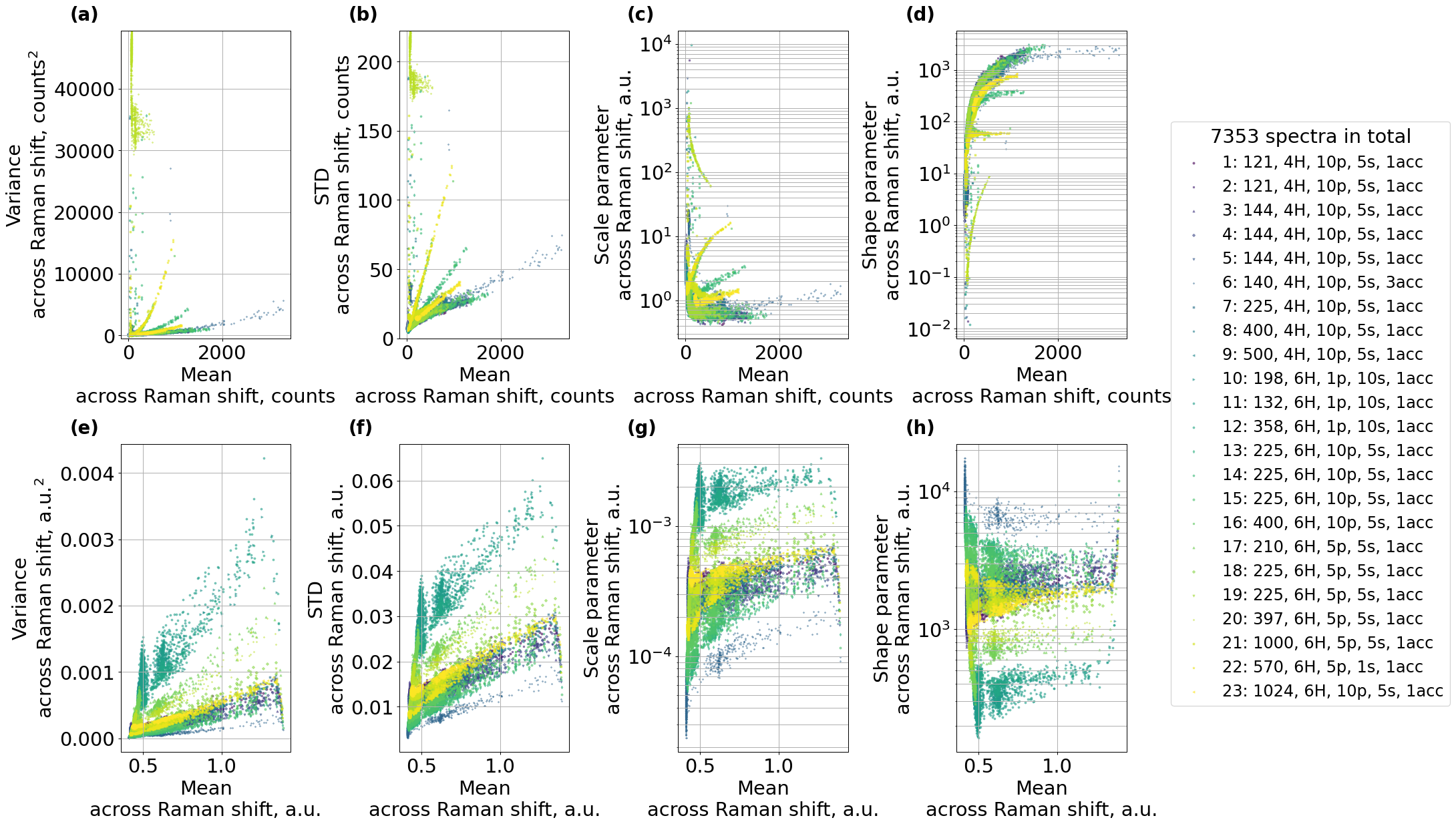}
    \caption{Statistical parameters of all the datasets. Different statistical parameters, such as mean, variance, standard deviation (STD), and scale and shape parameters - specific Gamma distribution parameters, for raw acquired data ( (a), (b), (c), (d) ) and for the same data being preprocessed for the model to be trained on ( (e), (f), (g), (h) ). Color of points represent different datafiles. Legend: dataset number, number of spectra, type of crystallinity, laser power [\%], acquisition time [s], number of accumulations.}
    \label{fig:stats}
\end{figure}

The Gamma distribution function choice is proven by Fig.~\ref{fig:stats}. Here, for each 2-dimensional dataset with dimensions \textbf{spectra} and \textbf{wavenumber}, across dataset slices at each wavenumber value, mean and variance values were calculated and plotted one versus another. Higher mean values correspond to the main SiC peak at 1525 cm$^{-1}$, while smaller -- for noise. It is expected, that the process of photon detection in Raman spectroscopy measurement is indeed Poissonian. However, it is not the case, since the main property of Poisson distribution \(variance = mean\) is not valid. There is not a linear, but quadratic dependence. This leads us to conclude that this particular registration process of random events occurring in a fixed interval of time should be still within a family of exponential probability functions, but more sophisticated. And since there is a dependence \(variance \propto mean^2\), which is a property of Gamma distribution, we deal with this type of probability function. Further investigations on what could be the reason of such behavior led us to the detector itself, which, according to its documentation, is following exactly Gamma distribution.

Loss function definition starts from Gamma distribution PDF:

\begin{equation}
    f(x)=\frac{1}{\Gamma(\alpha)\theta^{\alpha}}x^{\alpha-1}e^{-x/\theta},
    \label{eq:gamma_pdf}
\end{equation}

where $\alpha$ and $\theta$ are shape and scale parameters, which define mean and variance of this distribution:

\begin{equation}
    \operatorname{mean}=\mu=\alpha\cdot\theta, \qquad \operatorname{variance}=\sigma^2=\alpha\cdot\theta^2=\frac{1}{\alpha}\cdot\mu^2.
    \label{eq:mean_var}
\end{equation}

Mean value in this case can be called model's \textit{prediction} value, and intensity value at the given Raman shift can be called \textit{true} value.

Next step is to rewrite Eq. S\ref{eq:gamma_pdf} in terms of its mean value and find its negative log-likelihood (NLL), also neglecting all constant values:

\begin{equation}
    \operatorname{NLL} \left[ f(x)\right] = \ln(\mu)+\frac{x}{\mu}.
    \label{eq:gamma_nll}
\end{equation}

Since it is required for a loss function to find its minimum at zero (loss is a non-negative number by design), one needs to subtract it, resulting with same form of Eq. S\ref{eq:gamma_loss}:

\begin{equation}
    L\left(x,\mu\right)=L\left(true,prediction\right)=\left( \frac{x}{\mu} -1 \right) - \left( \ln\left(\frac{x}{\mu}\right) \right)
\end{equation}

Data after preprocessing kept its statistical properties (Fig.~\ref{fig:stats}). Preprocessing also helped to make Gamma distribution parameters more uniform both across different datafiles, reducing span of parameters on Y axis, but also within a datafile, flattening them out (Fig.~\ref{fig:stats} (g), (h)). 

\begin{figure}[!ht]
    \centering
    \includegraphics[width=0.85\linewidth]{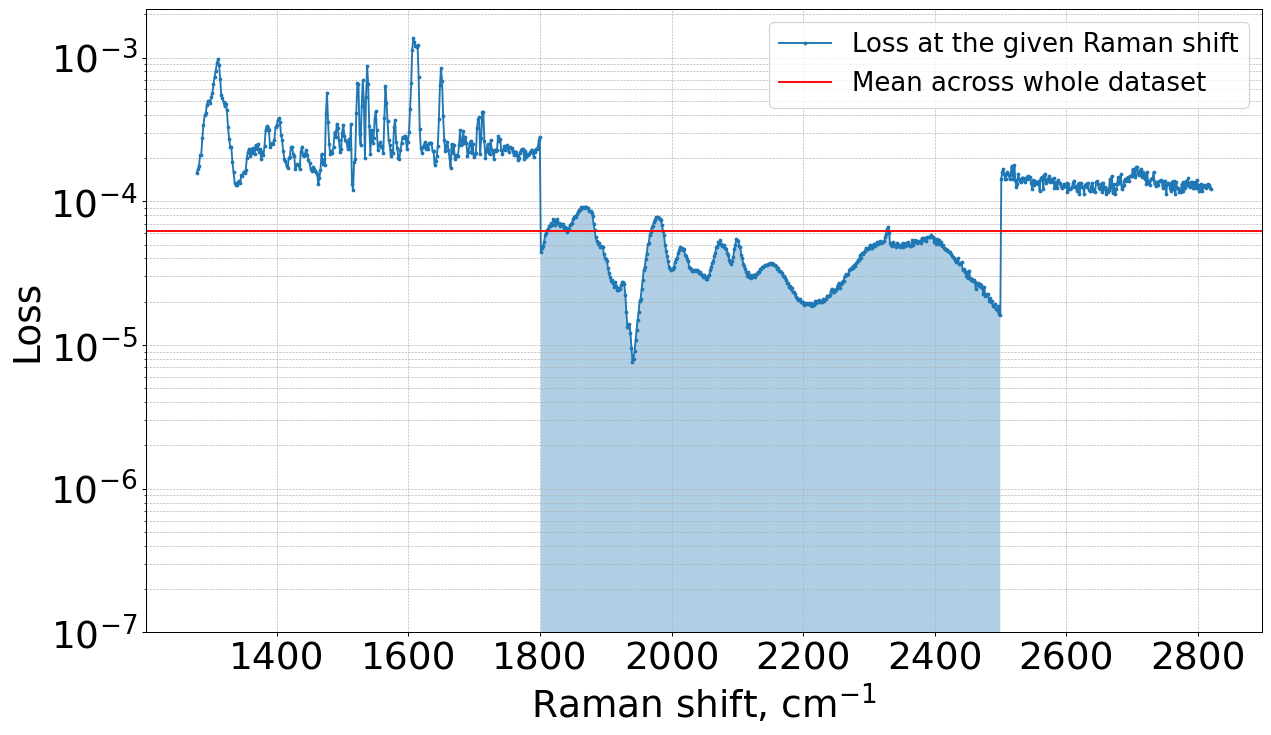}
    \caption{Loss calculation process visualization. Data values allowed to the model to be seen (unmasked) are highlighted by area filled regions.}
    \label{fig:loss_calc}
\end{figure}

An example of loss function values calculated for an individual spectrum is presented on Fig.S\ref{fig:loss_calc}. Each value is obtained by applying Eq. S\ref{eq:gamma_loss} on corresponding values of original SiC spectrum (true) and generated SiC spectrum (prediction). The closer the prediction to the true values - the smaller the loss. It can be clearly seen that for unmasked data loss is $\sim$ 2 o.o.m. lower than for hidden regions, implying effectively reproducing mechanism for a given part of a sequence (copy-pasting).

\begin{figure}[!ht]
    \centering
    \includegraphics[width=0.85\linewidth]{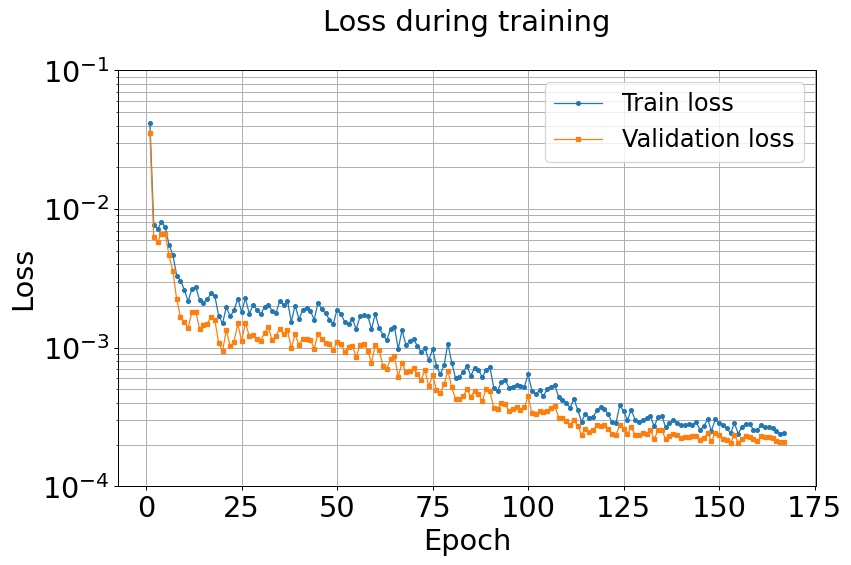}
    \caption{Model loss throughout the training process.}
    \label{fig:loss}
\end{figure}

\begin{figure}[!ht]%
    \centering
    \includegraphics[width=0.75\linewidth]{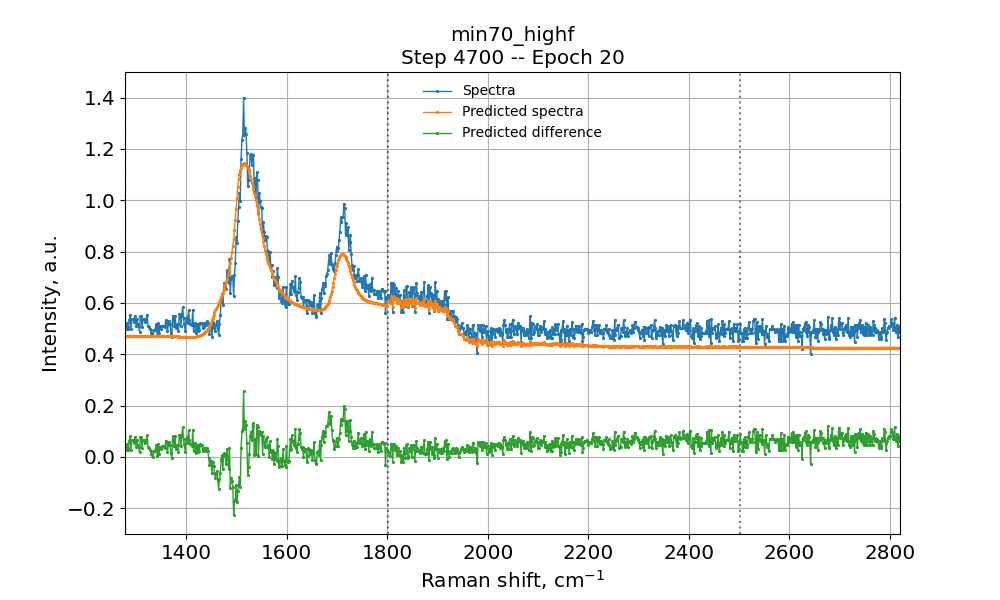}
    \includegraphics[width=0.75\linewidth]{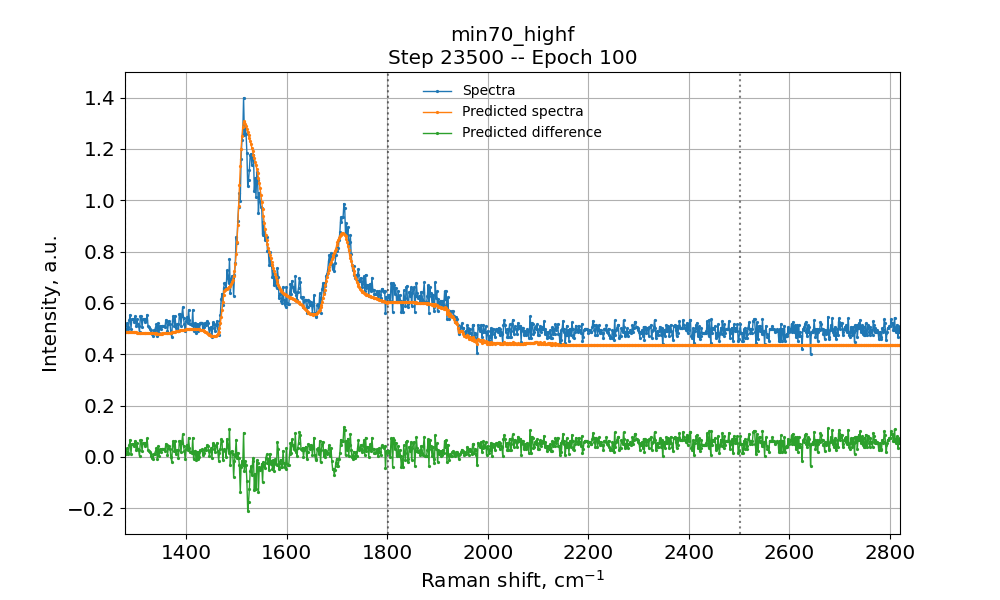}
    \includegraphics[width=0.75\linewidth]{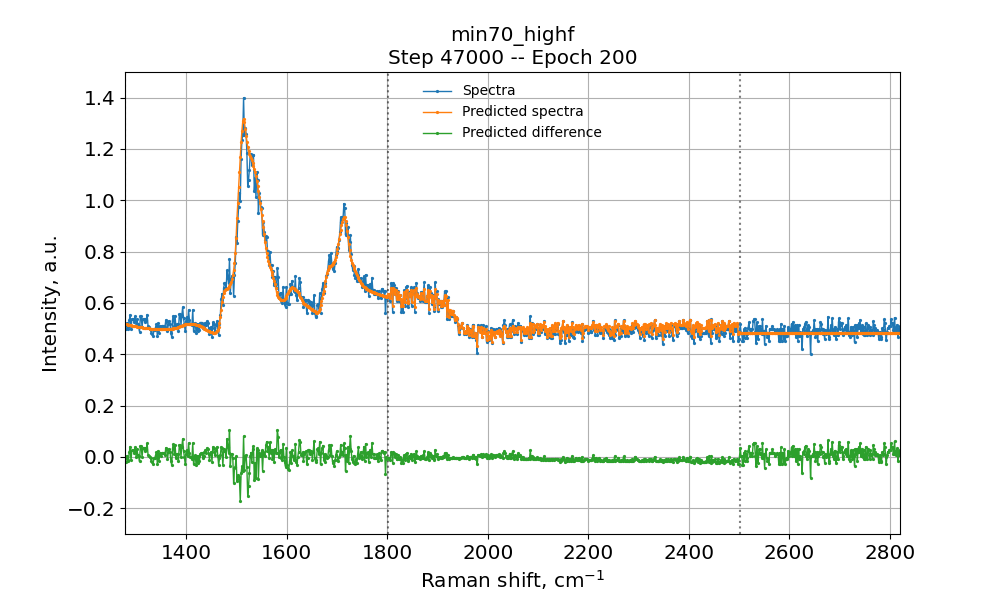}
    \caption{Model predictions captured at intermediate stages of training process: after epoch 20, epoch 100 and epoch 200.}
    \label{fig:pred:main}
\end{figure}

On the loss versus epoch curves (Fig.~\ref{fig:loss}) it is clearly seen that 1) no overfitting was reached (train loss is not continuously decreasing and validation loss is not stuck in a certain value, i.e. no line crossing reached), 2) there are several plateaus across the training process. The latter is explained as the following by looking on the model prediction output throughout the training (Fig.~\ref{fig:pred:main}): the model has successfully captured the main SiC peak located at 1525 cm$^{-1}$ after 20 epochs; then in-between epochs 20 and 100 it was trying its hypotheses about minor features, capturing the correct way of weights adjustment up to epoch 200; then, the rest of training was a fine-tuning to squeeze the most similarity out from the model predictions compared to the original dataset, reaching another local minimum on loss hyperplane. The training was explicitly stopped right before overfitting criteria.

\subsection{Model architecture parameters}

During several training processes optimal model architecture parameters were found (Table S\ref{table:model_parameters}). Since the task is not at the Natural Language Processing (NLP) level, we have rapidly reduced common transformer model's parameters (e.g. for embedding dimension: from several thousands to 64). It is also convenient to use powers of 2 in the model parameters values for easier computation.

\begin{center}
\begin{table}[!ht]
        \centering
        \begin{tabular}{|c|c|c|c|c|c|}
            \hline
            Parameter   & \makecell{Embedding\\dimension}   & Heads & Layers    & \makecell{Dropout\\rate}   & \makecell{Masked regions\\(hidden from the model)}    \\
            \hline
            Value       &               64                  &  8    &  2        &  0.2      &       \begin{tabular}{ r c c c l }
                ( & -1 & -- & 1800 & ) \\
                ( & 2500 & -- & 9999 & )
            \end{tabular} \\
            \hline
        \end{tabular}
\caption{Optimal model architecture parameters found}
\label{table:model_parameters}
\end{table}
\end{center}

\stepcounter{section}%
\section*{Supplementary Note \thesection: Model performance evaluation}

\begin{figure}[!ht]
    \centering
    \includegraphics[width=0.5\linewidth]{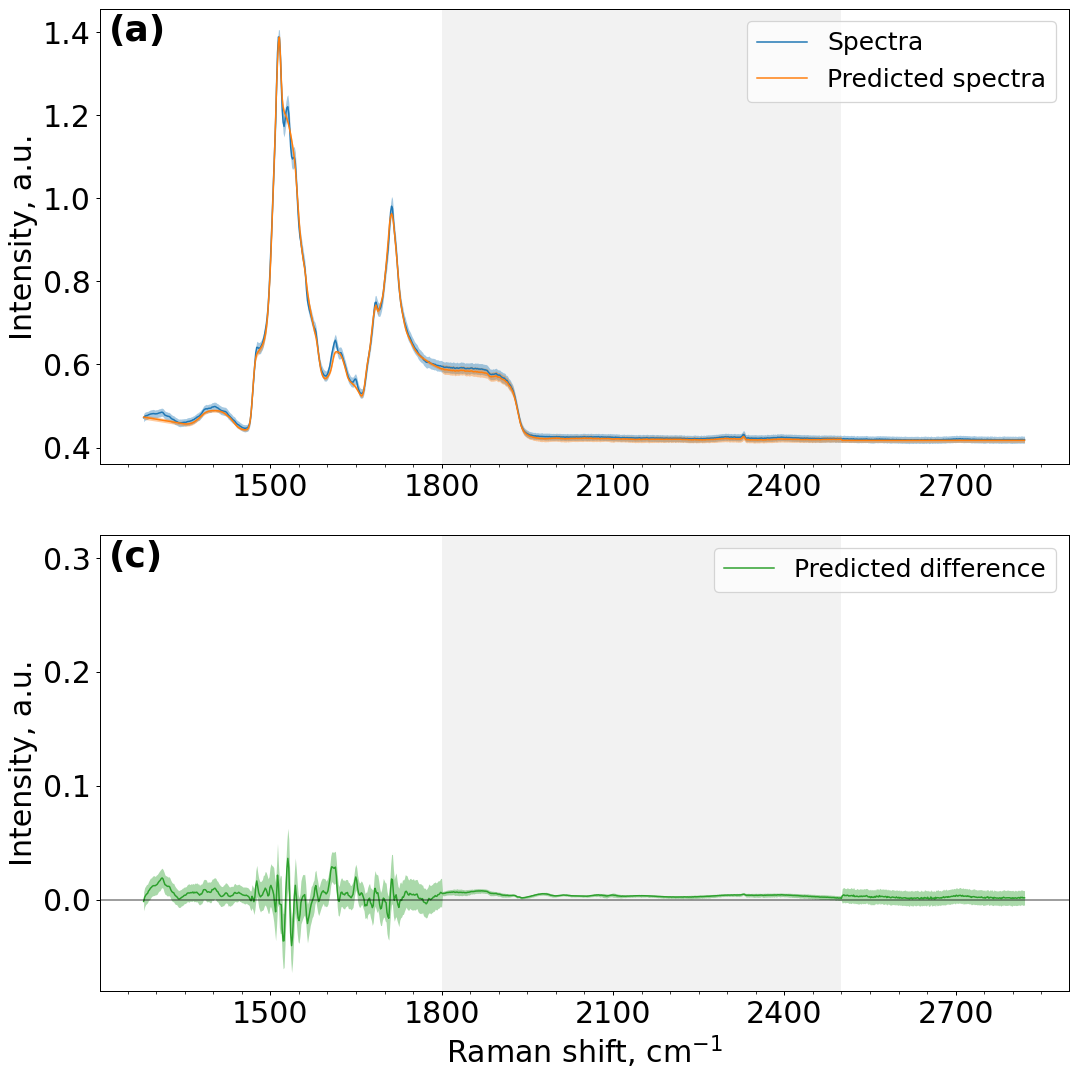}\includegraphics[width=0.5\linewidth]{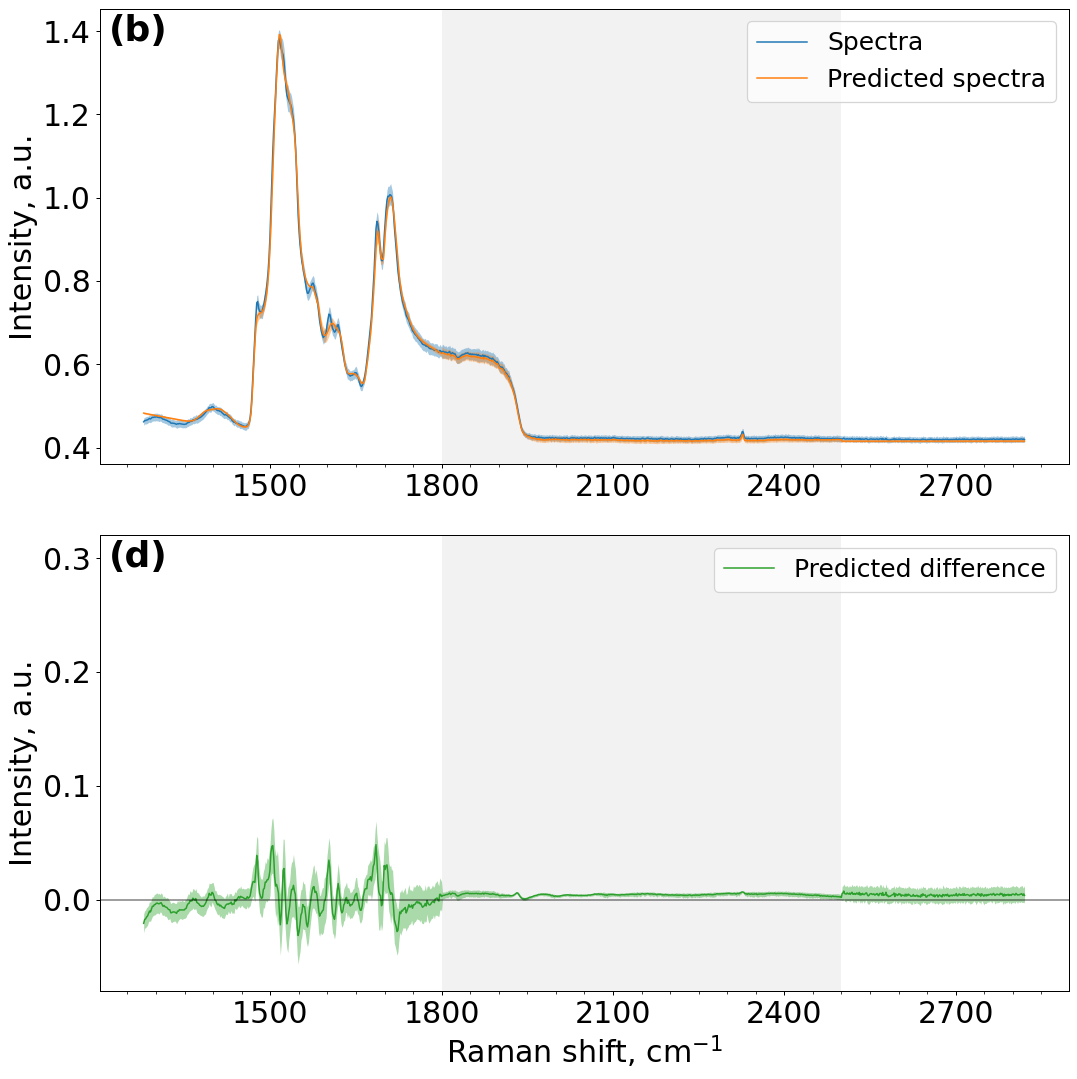}
    \caption{Model output after the training with different inputs: (a,c) 6H-SiC Raman spectrum is given that was a part of training dataset and (b,d) 4H-SiC Raman spectrum is given that was not a part of training dataset, allowing to estimate the accuracy of the model by subtraction of generated SiC spectrum; gray shaded region is the region of data available to the model.}
    \label{fig:SiC-pred}
\end{figure}

We can visualize the model performance by providing to it SiC spectra and looking at subtraction residuals with generated SiC spectra (Fig.~\ref{fig:SiC-pred}): they are centered around close-to-zero value with small variance. It is noticeable that noisiest parts of residuals are hidden masked regions. This is due to the internal transformer's mechanism: with a text example in NLP task, it is not only providing the next word in a sequence, but the full sequence, effectively copy-pasting known part; in the case of spectral data, it generates datapoints in such a way of effectively repeating shown data, while providing the most probable value in the region of interest.

Measurement conditions: 100x objective magnification, 532 nm green laser, 1800 lines/mm grating, 5 s acquisition time, 5\% laser power, and number of accumulations equal to 1.

\stepcounter{section}%
\section*{Supplementary Note \thesection: \textit{Ab initio} calculations}

\subsection{Choice of active atoms for frozen-phonon displacements}

\begin{figure}[!ht]
  \centering
  \includegraphics[width=0.85\textwidth]{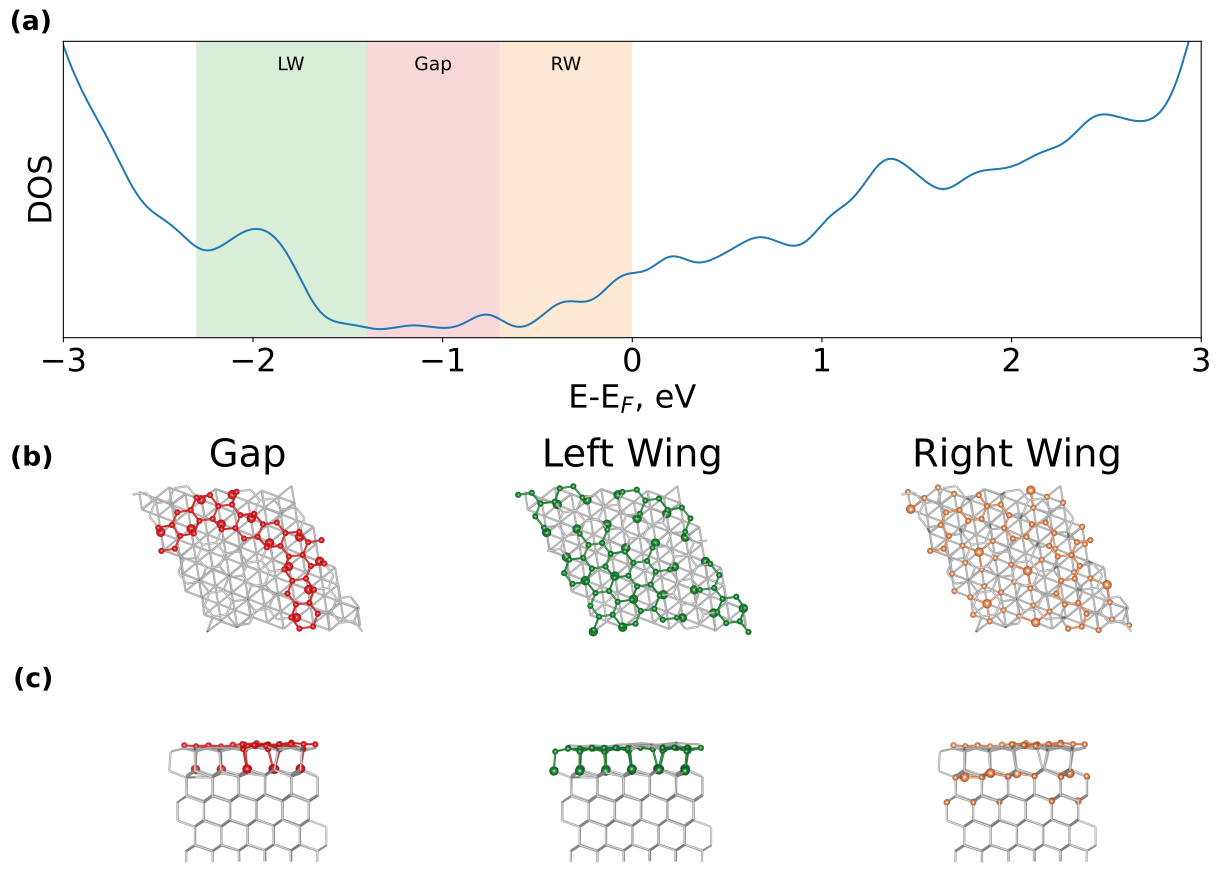} 
  \caption{(a) Total DOS around the Fermi level ($E=0$), partitioned into the three energy windows used to define the state-resolved active sets: left wing (LW), gap, and right wing (RW) states. (b) Top views and (c) side views of the atoms selected for frozen-phonon displacements for each window (colored), shown on top of the full structure (gray).}
  \label{fig:theory_S1}
\end{figure}

In the frozen-phonon approach, interatomic force constants are evaluated from finite differences of forces upon small atomic displacements. For the ZLG, the electronic states around the Fermi level are strongly inhomogeneous in real space and correlate with well-defined structural motifs (crests vs. tiles) and bonding patterns at the interface. In particular, Ref.~\cite{Cavallucci2018} showed that: the in-gap states are localized on the crests and involve the intruding (Si-C bonded) sites, the states just below the gap (left wing, LW) are largely associated with the ``benzene-like'' units inside the tiles and with the bonded interstitial sites, and the states above the gap up to $E_F$ (right wing, RW) have a pronounced subsurface component related to Si dangling bonds beneath the tiles, with an additional surface contribution on the crests. 
This localization allows restricting the atomic displacements to a subset of active atoms that (a) carry most of the localized density of states (LDOS) in a given energy window and (b) include the first-bonded partners required to preserve the relevant interfacial bonding/dangling-bond motifs.  
We partition the occupied states near-$E_F$ into three windows (Fig.~\ref{fig:theory_S1}a), consistently with the decomposition used in Ref.~\cite{Cavallucci2018}:
\begin{align*}
\mathrm{LW}:&\quad E-E_F \in [-2.3,-1.4]~\mathrm{eV}, \nonumber\\
\mathrm{Gap}:&\quad E-E_F \in [-1.4,-0.7]~\mathrm{eV}, \nonumber\\
\mathrm{RW}:&\quad E-E_F \in [-0.7,0]~\mathrm{eV}. \label{eq:windows}
\end{align*}
For each window $I \in \{\mathrm{LW},\mathrm{Gap},\mathrm{RW}\}$, we identify the atoms that spatially overlap with the LDOS isosurface. All atoms not belonging to the active set for a given window are kept fixed at the relaxed geometry during the frozen-phonon finite-difference steps, allowing for a consistent reduction of the computational workload.

The resulting active atoms are shown in Fig.~\ref{fig:theory_S1}b,c and define three window-dependent sets:

\begin{itemize}
  \item \textbf{Gap states}
  The in-gap states are concentrated on the crest network, with an additional inner component associated with the intruding crest sites that are covalently bound to the substrate.
  Accordingly, we displaced the crest C atoms (protruding network) and the directly connected interface partners, i.e. the intruding ZLG C atoms participating in Si-C bonds and the corresponding topmost substrate Si atoms.
  This choice targets the bonding nature of the in-gap states (between crest atoms and/or between intruding crest atoms and the substrate) \cite{Cavallucci2018}.

  \item \textbf{Left wing states}
  The LW states are primarily associated with the benzene-like rings within the tiles, while retaining a subsurface component underneath the tiles on bonded interstitial sites.
  In this case, we displaced the C atoms forming the benzene-like units, the nearby bonded interstitial ZLG C atoms, and their Si partners at the interface. 

  \item \textbf{Right wing states}
  Finally, the RW states feature a strong subsurface contribution localized below the tiles attributed to Si dangling bonds (with a surface component resembling localized $p_z$ orbitals on/near the crests) \cite{Cavallucci2018}.  For this reason, we included the substrate atoms carrying the dangling-bond LDOS under the selected tiles (and their nearest neighbors in the first few SiC layers), together with the subset of nearby crest atoms that overlap with the surface RW LDOS component.
\end{itemize}

\subsection{Vibrational density of states and state-resolved projections}

\begin{figure}[!ht]
  \centering
  \includegraphics[width=0.85\textwidth]{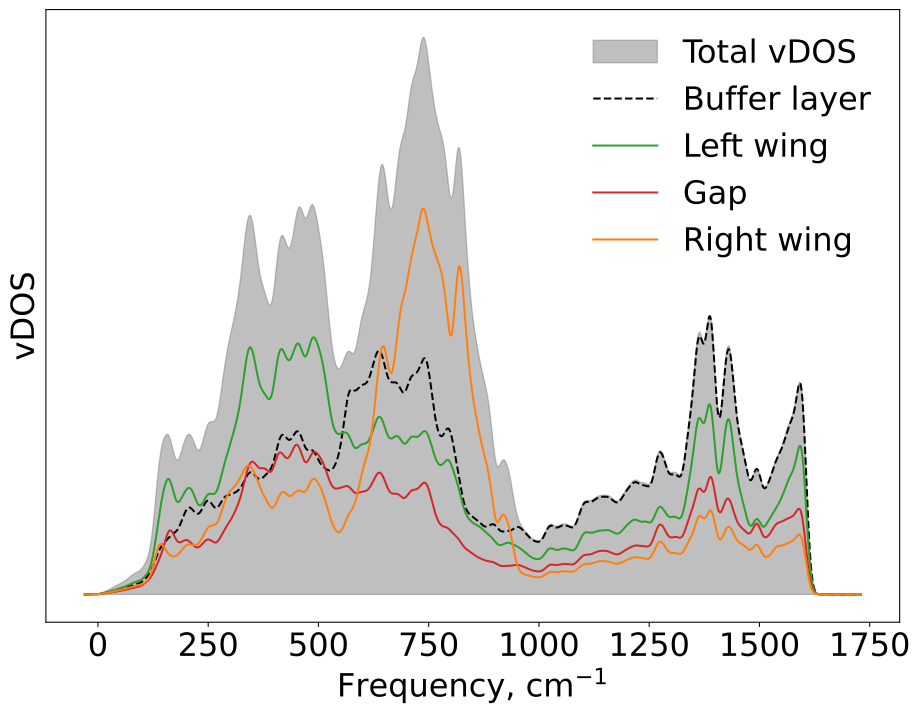}
  \caption{Total vDOS of the ZLG/SiC slab (gray) and projected vDOS onto the full ZLG (black dashed) and onto the LW/Gap/RW state-selected atomic subsets (green/red/orange; see Fig.~\ref{fig:theory_S1} for definitions).}
  \label{fig:theory_S2}
\end{figure}

Fig.~\ref{fig:theory_S2} reports the vibrational density of states (vDOS) of the full ZLG/SiC slab (gray shaded area), together with projections onto the ZLG atoms (black dashed line) and onto the three atomic subsets defined above (colored lines: LW, Gap, RW). In the high-frequency region (upper band above $\sim$1000 cm$^{-1}$ and up to $\sim$1600 cm$^{-1}$), the vDOS intensity is almost completely determined by ZLG atoms: the highest-frequency vibrations are largely associated with C-C bond-stretching within the ZLG network, while the substrate increases the spectral weight at low and intermediate frequencies.

\subsection{Raman activity and atomic decomposition}

\begin{figure}[!ht]
  \centering
  \includegraphics[width=0.85\textwidth]{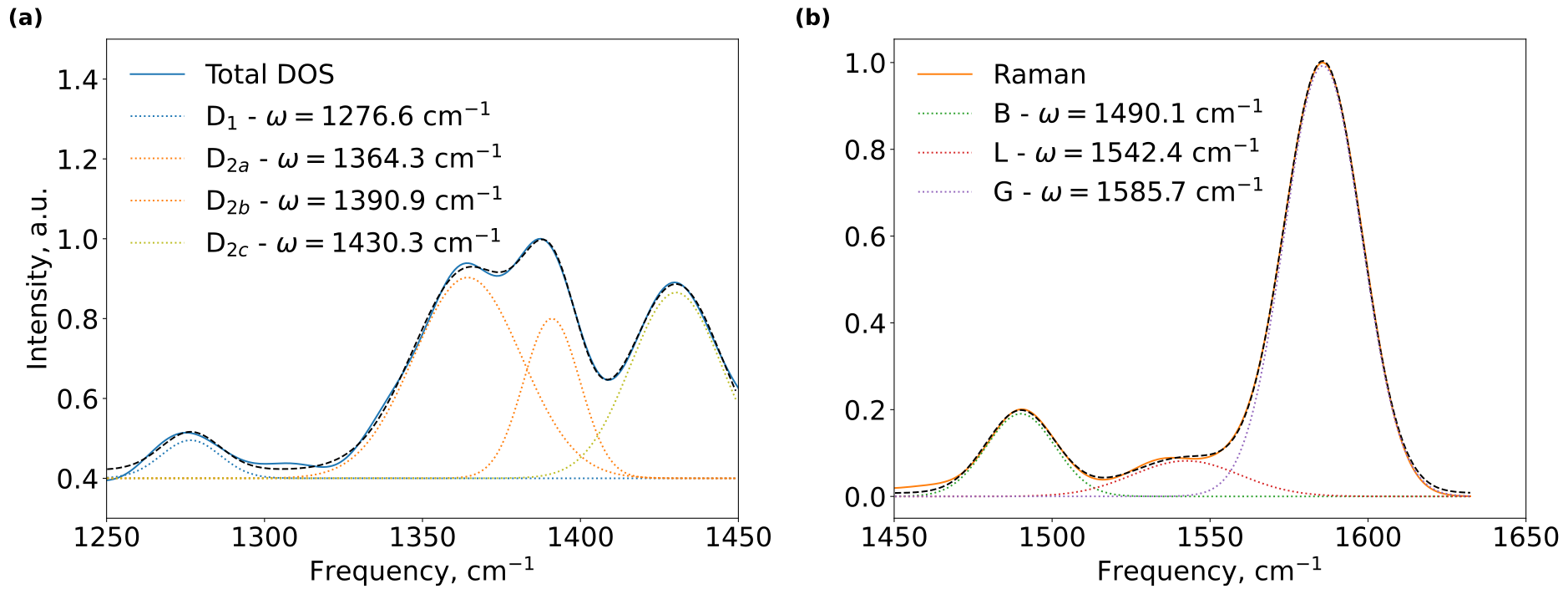}
  \caption{(a) vDOS-based decomposition of the D region into the individual contributions ($D_1$, $D_{2a}$, $D_{2b}$, $D_{2c}$). (b) Raman-proxy spectrum (orange) and the fit of the B, L, and G components, with peak positions indicated in the legend.}
  \label{fig:theory_S3}
\end{figure}

To highlight the ZLG vibrations that are most likely to contribute to the first-order Raman response, we construct a Raman-like proxy by projecting each $\Gamma$-point ZLG normal mode onto the Raman-active graphene $E_{2g}$ displacement pattern (mapped onto the ZLG in-plane frame), following a similar approach adopted by Radtke and Lazzeri \cite{Radtke2025}.
For a ZLG mode $\nu$ with mass-normalized eigenvector $e^{(\nu)}_{I\alpha}$, the proxy weight is
\begin{equation}
W_\nu =\left|\sum_{I\in \mathrm{ZLG}}\sum_{\alpha=x,y}\, e^{(\nu)}_{I\alpha}\, e^{(E_{2g})}_{I\alpha}\right|^{2}
\label{eq:raman_proxy}
\end{equation}
where both eigenvectors are normalized such that $\sum_{I,\alpha}|e_{I\alpha}|^{2}=1$.
A continuous Raman-proxy spectrum is obtained by broadening the discrete phonon lines:
\begin{equation}
I_{\mathrm{Raman}}(\omega) = \sum_{\nu} W_\nu \, G(\omega-\omega_\nu),
\label{eq:raman_proxy_spectrum}
\end{equation}
with $G$ the chosen broadening function and $\omega_\nu$ the phonon frequencies. In what follows, we adopted a Gaussian broadening function with $\sigma=10$ cm$^{-1}$. Fig.~\ref{fig:theory_S3}b reports Raman-like intensity (orange) together with a fit of the three main components (B, L, and G).

To localize the origin of a given spectral feature, we define a per-atom contribution for each mode,
\begin{equation}
w_{I}^{(\nu)} =\sum_{\alpha=x,y}\left|\,e^{(\nu)}_{I\alpha}\, e^{(E_{2g})}_{I\alpha}\right|^{2},
\qquad
\sum_{I\in\mathrm{ZLG}} w_{I}^{(\nu)} = W_\nu ,
\label{eq:atomic_decomp}
\end{equation}
and sum $w_{I}^{(\nu)}$ over the modes belonging to a given frequency interval (e.g.\ B, L, or G) to obtain the atom-resolved intensity maps shown in the main text, Fig.~4b,c.

\subsection{Atomic decomposition of the D-band features}

The Raman-proxy construction does not yield any intensity in the D-frequency region. We fitted four different peak contributions in the vDOS, as shown in Fig.~\ref{fig:theory_S3}a. Specifically, in the range 1250-1450~cm$^{-1}$ we identify four peaks ($D_1$, $D_{2a}$, $D_{2b}$, $D_{2c}$). For each component $D_i$, we then compute an atom-resolved contribution by summing the normalized eigenvector weight of the corresponding modes on each atom,
\begin{equation}
A_I^{(D_i)} = \sum_{\nu \in D_i}\sum_{\alpha=x,y,z}\left|e^{(\nu)}_{I\alpha}\right|^2,
\label{eq:D_atomic_weight}
\end{equation}
where the sum runs over the set of modes $\nu$ assigned to the peak $D_i$ (within the frequency window of that component) and $e^{(\nu)}_{I\alpha}$ are mass-normalized phonon eigenvectors.

\begin{figure}[!ht]
  \centering
  \includegraphics[width=0.85\textwidth]{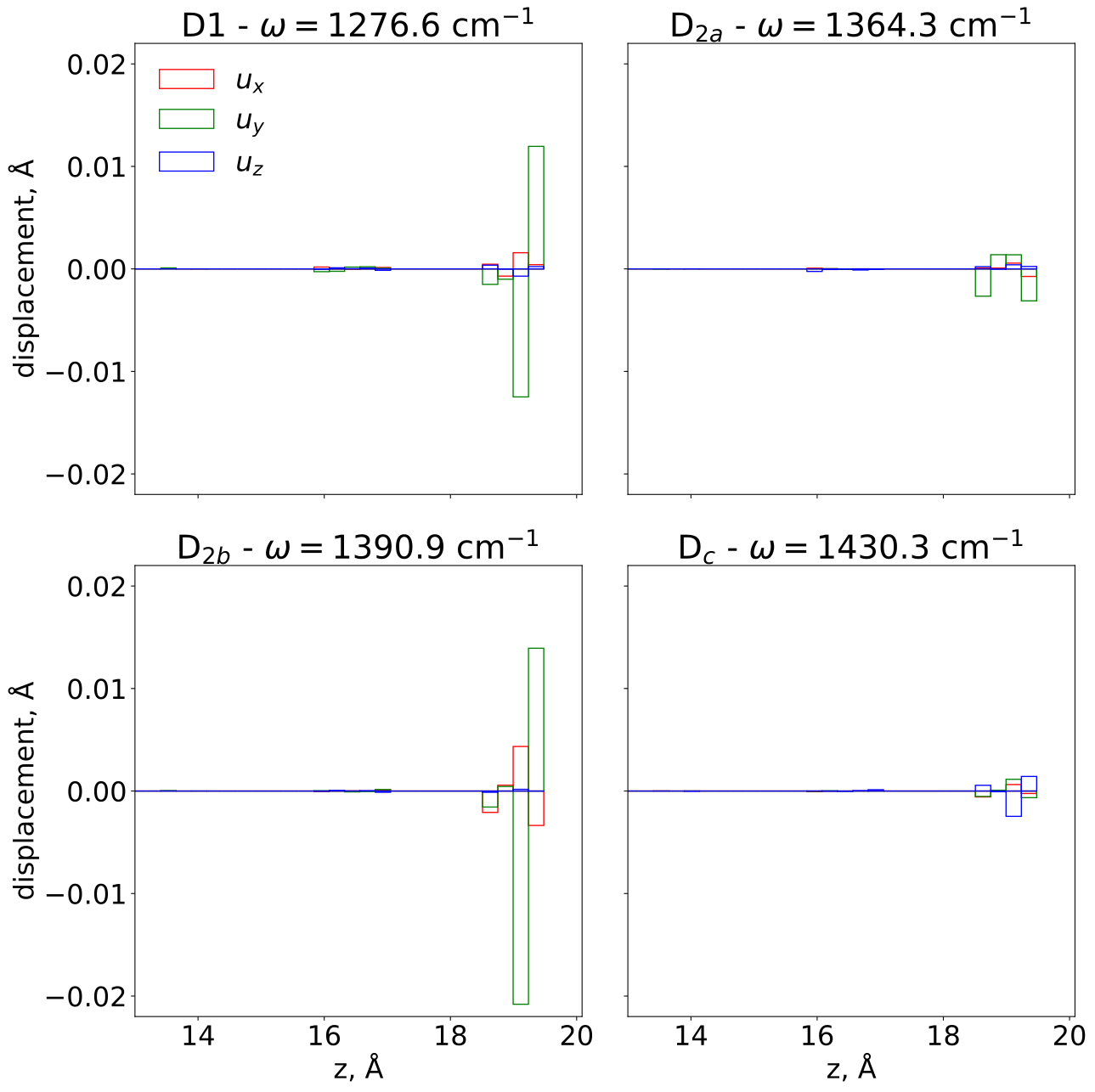}
  \caption{Displacement profiles in the three directions $x,y,z$ along the out-of-plane coordinates $z$ for the ZLG/SiC atoms associated with the four D-band components identified in Fig.~\ref{fig:theory_S3}a. For each peak ($D_1$, $D_{2a}$, $D_{2b}$, $D_{2c}$), the Cartesian displacement components ($u_x$, $u_y$, $u_z$) are shown as a function of atomic height $z$, highlighting that the vibrational motion is strongly localized within the buffer-layer region.}
  \label{fig:theory_S4}
\end{figure}

Although the D-band decomposition is performed on the \emph{total} vDOS spectrum, our analysis focuses only on ZLG atoms. In fact, the displacement patterns of the atoms associated with each extracted peak (Fig.~\ref{fig:theory_S4}) show that the corresponding mode displacements are strongly localized within the ZLG region, for all four peaks. This real-space localization is consistent with the projected vDOS analysis showed above in Fig.~\ref{fig:theory_S2}, in which the ZLG contribution remains predominantly within the D-frequency range, indicating that these modes are predominantly carried by the reconstructed ZLG rather than by bulk-like SiC vibrations. 

\subsection{Morphological descriptors of the ZLG}

\begin{figure}[!ht]
  \centering
  \includegraphics[width=0.85\textwidth]{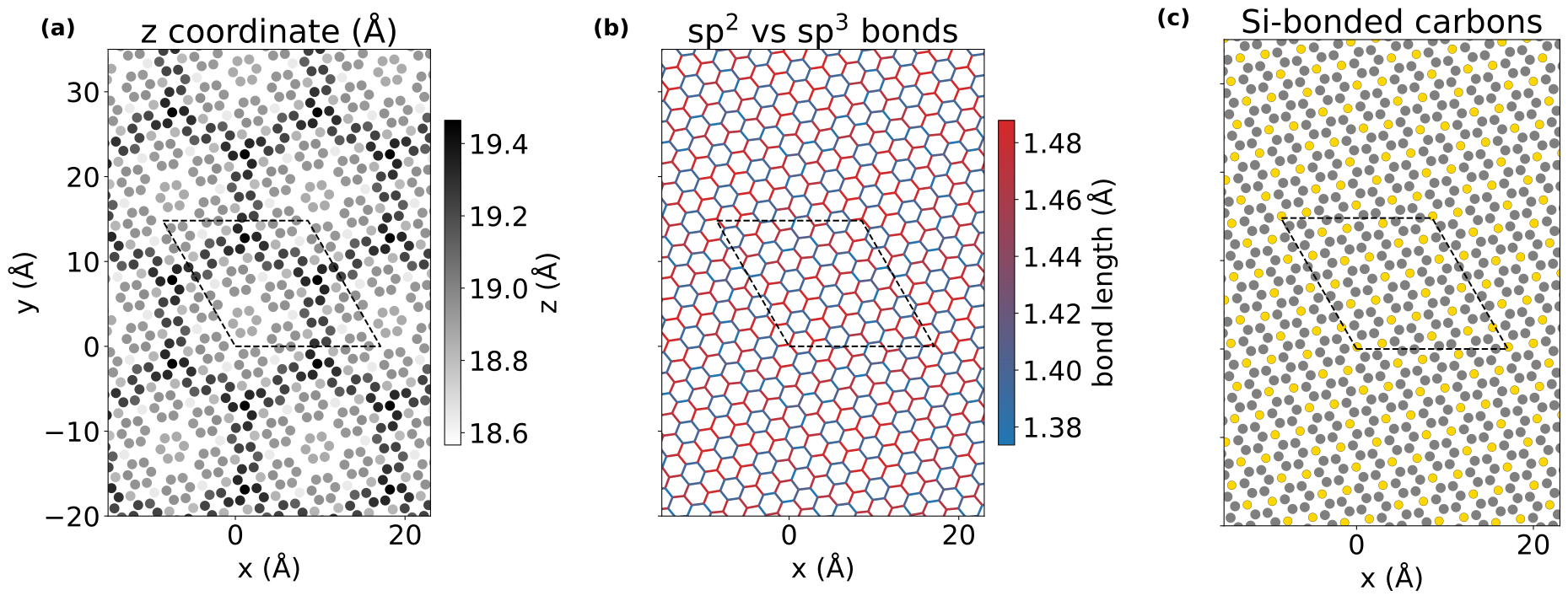}
  \caption{Structural/morphological descriptors of the ZLG. (a) In-plane map of the atomic height ($z$ coordinate). (b) Classification of C-C bonds into $sp^{2}$-like and $sp^{3}$-like based on bond length, with characteristic centers at $d\approx1.40~\text{\AA}$ ($sp^{2}$) and $d\approx1.47~\text{\AA}$ ($sp^{3}$). (c) Identification of Si-bonded ZLG carbon atoms (gold) versus non-bonded ZLG carbons (gray). The dashed parallelogram indicates the ZLG supercell.}
  \label{fig:theory_S5}
\end{figure}

Fig.~\ref{fig:theory_S5} summarizes the main real-space descriptors used to characterize the reconstructed ZLG and to correlate vibrational/spectral features with local structure.
Fig.~\ref{fig:theory_S5}a reports the ZLG height field, shown as the atomic $z$ coordinate mapped onto the in-plane $(x,y)$ positions. The distribution highlights the intrinsic ZLG corrugation, with elevated crest regions and lower tile interiors arranged in a superlattice pattern (dashed parallelogram indicates the ZLG supercell).

Fig.~\ref{fig:theory_S5}b classifies C-C bonds based on their bond length, providing a structural aid to distinguish more graphene-like ($sp^{2}$) from more tetrahedral-like ($sp^{3}$) environments within the ZLG. We find a bimodal bond-length distribution: the $sp^{2}$-like bonds are centered around $d \simeq 1.40~\text{\AA}$, while the $sp^{3}$-like bonds peak around $d \simeq 1.47~\text{\AA}$. This separation allows us to identify the two contributions across the ZLG network.

Finally, Fig.~\ref{fig:theory_S5}c shows the subset of ZLG carbon atoms that form covalent bonds with the SiC substrate. Si-bonded carbon atoms are highlighted in gold, whereas non-bonded ZLG carbon atoms are shown in gray. This representation directly identifies the interfacial anchoring sites (typically associated with intruding atoms and locally $sp^{3}$-like environments), which are crucial for interpreting both the structural reconstruction and the localization of vibrational/electronic features discussed in the main text.

\pagebreak
\pagebreak
\printbibliography[title={Supplementary Information References}]

@article{ghosh_hyperspectral_2022,
	title = {Hyperspectral Unmixing Using Transformer Network},
	volume = {60},
	issn = {1558-0644},
	url = {https://ieeexplore.ieee.org/document/9848995},
	doi = {10.1109/TGRS.2022.3196057},
	pages = {1--16},
	journaltitle = {{IEEE} Transactions on Geoscience and Remote Sensing},
	author = {Ghosh, Preetam and Roy, Swalpa Kumar and Koirala, Bikram and Rasti, Behnood and Scheunders, Paul},
	urldate = {2026-01-06},
	date = {2022},
}

@article{wang_raman_2019,
	title = {Raman Scattering Study of Silicon Carbide Irradiated with 1.25 {MeV} Si Ions},
	volume = {493},
	issn = {1757-899X},
	url = {https://doi.org/10.1088/1757-899X/493/1/012092},
	doi = {10.1088/1757-899X/493/1/012092},
	pages = {012092},
	number = {1},
	journaltitle = {{IOP} Conference Series: Materials Science and Engineering},
	shortjournal = {{IOP} Conf. Ser.: Mater. Sci. Eng.},
	author = {Wang, Pengfei and Wang, Shuai},
	urldate = {2026-01-06},
	date = {2019-03},
	langid = {english},
	note = {Publisher: {IOP} Publishing},
}

@article{van_bommel_leed_1975,
	title = {{LEED} and Auger electron observations of the {SiC}(0001) surface},
	volume = {48},
	issn = {0039-6028},
	url = {https://www.sciencedirect.com/science/article/pii/0039602875904197},
	doi = {10.1016/0039-6028(75)90419-7},
	pages = {463--472},
	number = {2},
	journaltitle = {Surface Science},
	shortjournal = {Surface Science},
	author = {Van Bommel, A. J. and Crombeen, J. E. and Van Tooren, A.},
	urldate = {2026-01-06},
	date = {1975-03-02},
}

@article{briggs_atomically_2020,
	title = {Atomically thin half-van der Waals metals enabled by confinement heteroepitaxy},
	volume = {19},
	rights = {2020 The Author(s), under exclusive licence to Springer Nature Limited},
	issn = {1476-4660},
	url = {https://www.nature.com/articles/s41563-020-0631-x},
	doi = {10.1038/s41563-020-0631-x},
	pages = {637--643},
	number = {6},
	journaltitle = {Nature Materials},
	shortjournal = {Nat. Mater.},
	author = {Briggs, Natalie and Bersch, Brian and Wang, Yuanxi and Jiang, Jue and Koch, Roland J. and Nayir, Nadire and Wang, Ke and Kolmer, Marek and Ko, Wonhee and De La Fuente Duran, Ana and Subramanian, Shruti and Dong, Chengye and Shallenberger, Jeffrey and Fu, Mingming and Zou, Qiang and Chuang, Ya-Wen and Gai, Zheng and Li, An-Ping and Bostwick, Aaron and Jozwiak, Chris and Chang, Cui-Zu and Rotenberg, Eli and Zhu, Jun and van Duin, Adri C. T. and Crespi, Vincent and Robinson, Joshua A.},
	urldate = {2026-01-05},
	date = {2020-06},
	langid = {english},
	note = {Publisher: Nature Publishing Group},
}

@article{castro_neto_electronic_2009,
	title = {The electronic properties of graphene},
	volume = {81},
	url = {https://link.aps.org/doi/10.1103/RevModPhys.81.109},
	doi = {10.1103/RevModPhys.81.109},
	pages = {109--162},
	number = {1},
	journaltitle = {Reviews of Modern Physics},
	shortjournal = {Rev. Mod. Phys.},
	author = {Castro Neto, A. H. and Guinea, F. and Peres, N. M. R. and Novoselov, K. S. and Geim, A. K.},
	urldate = {2026-01-05},
	date = {2009-01-14},
	note = {Publisher: American Physical Society},
}

@article{bruna_doping_2014,
	title = {Doping Dependence of the Raman Spectrum of Defected Graphene},
	volume = {8},
	issn = {1936-0851},
	url = {https://doi.org/10.1021/nn502676g},
	doi = {10.1021/nn502676g},
	pages = {7432--7441},
	number = {7},
	journaltitle = {{ACS} Nano},
	shortjournal = {{ACS} Nano},
	author = {Bruna, Matteo and Ott, Anna K. and Ijäs, Mari and Yoon, Duhee and Sassi, Ugo and Ferrari, Andrea C.},
	urldate = {2025-12-30},
	date = {2014-07-22},
	note = {Publisher: American Chemical Society},
}

@article{ferrari_raman_2013,
	title = {Raman spectroscopy as a versatile tool for studying the properties of graphene},
	volume = {8},
	rights = {2013 Springer Nature Limited},
	issn = {1748-3395},
	url = {https://www.nature.com/articles/nnano.2013.46},
	doi = {10.1038/nnano.2013.46},
	pages = {235--246},
	number = {4},
	journaltitle = {Nature Nanotechnology},
	shortjournal = {Nature Nanotech},
	author = {Ferrari, Andrea C. and Basko, Denis M.},
	urldate = {2025-12-09},
	date = {2013-04},
	langid = {english},
	note = {Publisher: Nature Publishing Group},
}

@article{forti_semiconductor_2020,
	title = {Semiconductor to metal transition in two-dimensional gold and its van der Waals heterostack with graphene},
	volume = {11},
	rights = {2020 The Author(s)},
	issn = {2041-1723},
	url = {https://www.nature.com/articles/s41467-020-15683-1},
	doi = {10.1038/s41467-020-15683-1},
	pages = {2236},
	number = {1},
	journaltitle = {Nature Communications},
	shortjournal = {Nat Commun},
	author = {Forti, Stiven and Link, Stefan and Stöhr, Alexander and Niu, Yuran and Zakharov, Alexei A. and Coletti, Camilla and Starke, Ulrich},
	urldate = {2025-12-09},
	date = {2020-05-06},
	langid = {english},
	note = {Publisher: Nature Publishing Group},
}

@article{wan_contrasting_2023,
	title = {Contrasting Transport Performance of Electron- and Hole-Doped Epitaxial Graphene for Quantum Resistance Metrology},
	volume = {40},
	issn = {0256-307X},
	url = {https://doi.org/10.1088/0256-307X/40/10/107201},
	doi = {10.1088/0256-307X/40/10/107201},
	pages = {107201},
	number = {10},
	journaltitle = {Chinese Physics Letters},
	shortjournal = {Chinese Phys. Lett.},
	author = {Wan, Xinyi and Fan, Xiaodong and Zhai, Changwei and Yang, Zhenyu and Hao, Lilong and Li, Lin and Lu, Yunfeng and Zeng, Changgan},
	urldate = {2025-12-09},
	date = {2023-10},
	langid = {english},
	note = {Publisher: Chinese Physical Society and {IOP} Publishing Ltd},
}

@article{janssen_quantum_2013,
	title = {Quantum resistance metrology using graphene},
	volume = {76},
	issn = {0034-4885},
	url = {https://doi.org/10.1088/0034-4885/76/10/104501},
	doi = {10.1088/0034-4885/76/10/104501},
	pages = {104501},
	number = {10},
	journaltitle = {Reports on Progress in Physics},
	shortjournal = {Rep. Prog. Phys.},
	author = {Janssen, T J B M and Tzalenchuk, A and Lara-Avila, S and Kubatkin, S and Fal'ko, V I},
	urldate = {2025-12-09},
	date = {2013-10},
	langid = {english},
	note = {Publisher: {IOP} Publishing},
}

@article{tzalenchuk_towards_2010,
	title = {Towards a quantum resistance standard based on epitaxial graphene},
	volume = {5},
	rights = {2010 Springer Nature Limited},
	issn = {1748-3395},
	url = {https://www.nature.com/articles/nnano.2009.474},
	doi = {10.1038/nnano.2009.474},
	pages = {186--189},
	number = {3},
	journaltitle = {Nature Nanotechnology},
	shortjournal = {Nature Nanotech},
	author = {Tzalenchuk, Alexander and Lara-Avila, Samuel and Kalaboukhov, Alexei and Paolillo, Sara and Syväjärvi, Mikael and Yakimova, Rositza and Kazakova, Olga and Janssen, T. J. B. M. and Fal'ko, Vladimir and Kubatkin, Sergey},
	urldate = {2025-12-09},
	date = {2010-03},
	langid = {english},
	note = {Publisher: Nature Publishing Group},
}

@article{cavallucci_intrinsic_2018,
	title = {Intrinsic structural and electronic properties of the Buffer Layer on Silicon Carbide unraveled by Density Functional Theory},
	volume = {8},
	rights = {2018 The Author(s)},
	issn = {2045-2322},
	url = {https://www.nature.com/articles/s41598-018-31490-7},
	doi = {10.1038/s41598-018-31490-7},
	pages = {13097},
	number = {1},
	journaltitle = {Scientific Reports},
	shortjournal = {Sci Rep},
	author = {Cavallucci, Tommaso and Tozzini, Valentina},
	urldate = {2025-12-09},
	date = {2018-08-30},
	langid = {english},
	note = {Publisher: Nature Publishing Group},
}

@article{goler_revealing_2013,
	title = {Revealing the atomic structure of the buffer layer between {SiC}(0   0   0   1) and epitaxial graphene},
	volume = {51},
	issn = {0008-6223},
	url = {https://www.sciencedirect.com/science/article/pii/S0008622312007026},
	doi = {10.1016/j.carbon.2012.08.050},
	pages = {249--254},
	journaltitle = {Carbon},
	shortjournal = {Carbon},
	author = {Goler, Sarah and Coletti, Camilla and Piazza, Vincenzo and Pingue, Pasqualantonio and Colangelo, Francesco and Pellegrini, Vittorio and Emtsev, Konstantin V. and Forti, Stiven and Starke, Ulrich and Beltram, Fabio and Heun, Stefan},
	urldate = {2025-12-09},
	date = {2013-01-01},
}

@article{emtsev_interaction_2008,
	title = {Interaction, growth, and ordering of epitaxial graphene on {SiC}\{0001\} surfaces: A comparative photoelectron spectroscopy study},
	volume = {77},
	url = {https://link.aps.org/doi/10.1103/PhysRevB.77.155303},
	doi = {10.1103/PhysRevB.77.155303},
	shorttitle = {Interaction, growth, and ordering of epitaxial graphene on {SiC}\{0001\} surfaces},
	pages = {155303},
	number = {15},
	journaltitle = {Physical Review B},
	shortjournal = {Phys. Rev. B},
	author = {Emtsev, K. V. and Speck, F. and Seyller, Th. and Ley, L. and Riley, J. D.},
	urldate = {2025-12-09},
	date = {2008-04-02},
	note = {Publisher: American Physical Society},
}

@article{emtsev_towards_2009,
	title = {Towards wafer-size graphene layers by atmospheric pressure graphitization of silicon carbide},
	volume = {8},
	rights = {2009 Springer Nature Limited},
	issn = {1476-4660},
	url = {https://www.nature.com/articles/nmat2382},
	doi = {10.1038/nmat2382},
	pages = {203--207},
	number = {3},
	journaltitle = {Nature Materials},
	shortjournal = {Nature Mater},
	author = {Emtsev, Konstantin V. and Bostwick, Aaron and Horn, Karsten and Jobst, Johannes and Kellogg, Gary L. and Ley, Lothar and {McChesney}, Jessica L. and Ohta, Taisuke and Reshanov, Sergey A. and Röhrl, Jonas and Rotenberg, Eli and Schmid, Andreas K. and Waldmann, Daniel and Weber, Heiko B. and Seyller, Thomas},
	urldate = {2025-12-09},
	date = {2009-03},
	langid = {english},
	note = {Publisher: Nature Publishing Group},
}

@article{gebeyehu_decoupled_2024,
	title = {Decoupled High-Mobility Graphene on Cu(111)/Sapphire via Chemical Vapor Deposition},
	volume = {36},
	rights = {© 2024 The Author(s). Advanced Materials published by Wiley-{VCH} {GmbH}},
	issn = {1521-4095},
	url = {https://onlinelibrary.wiley.com/doi/abs/10.1002/adma.202404590},
	doi = {10.1002/adma.202404590},
	pages = {2404590},
	number = {44},
	journaltitle = {Advanced Materials},
	author = {Gebeyehu, Zewdu M. and Mišeikis, Vaidotas and Forti, Stiven and Rossi, Antonio and Mishra, Neeraj and Boschi, Alex and Ivanov, Yurii P. and Martini, Leonardo and Ochapski, Michal W. and Piccinini, Giulia and Watanabe, Kenji and Taniguchi, Takashi and Divitini, Giorgio and Beltram, Fabio and Pezzini, Sergio and Coletti, Camilla},
	urldate = {2025-12-09},
	date = {2024},
	langid = {english},
	note = {\_eprint: https://advanced.onlinelibrary.wiley.com/doi/pdf/10.1002/adma.202404590},
}

@article{pezzini_high-quality_2020,
	title = {High-quality electrical transport using scalable {CVD} graphene},
	volume = {7},
	issn = {2053-1583},
	url = {https://doi.org/10.1088/2053-1583/aba645},
	doi = {10.1088/2053-1583/aba645},
	pages = {041003},
	number = {4},
	journaltitle = {2D Materials},
	shortjournal = {2D Mater.},
	author = {Pezzini, Sergio and Mišeikis, Vaidotas and Pace, Simona and Rossella, Francesco and Watanabe, Kenji and Taniguchi, Takashi and Coletti, Camilla},
	urldate = {2025-12-09},
	date = {2020-08},
	langid = {english},
	note = {Publisher: {IOP} Publishing},
}

@article{banszerus_ultrahigh-mobility_2015,
	title = {Ultrahigh-mobility graphene devices from chemical vapor deposition on reusable copper},
	volume = {1},
	issn = {2375-2548},
	doi = {10.1126/sciadv.1500222},
	pages = {e1500222},
	number = {6},
	journaltitle = {Science Advances},
	shortjournal = {Sci Adv},
	author = {Banszerus, Luca and Schmitz, Michael and Engels, Stephan and Dauber, Jan and Oellers, Martin and Haupt, Federica and Watanabe, Kenji and Taniguchi, Takashi and Beschoten, Bernd and Stampfer, Christoph},
	date = {2015-07},
	pmid = {26601221},
	pmcid = {PMC4646786},
}

@article{li_large-area_2011,
	title = {Large-Area Graphene Single Crystals Grown by Low-Pressure Chemical Vapor Deposition of Methane on Copper},
	volume = {133},
	issn = {0002-7863},
	url = {https://doi.org/10.1021/ja109793s},
	doi = {10.1021/ja109793s},
	pages = {2816--2819},
	number = {9},
	journaltitle = {Journal of the American Chemical Society},
	shortjournal = {J. Am. Chem. Soc.},
	author = {Li, Xuesong and Magnuson, Carl W. and Venugopal, Archana and Tromp, Rudolf M. and Hannon, James B. and Vogel, Eric M. and Colombo, Luigi and Ruoff, Rodney S.},
	urldate = {2025-12-09},
	date = {2011-03-09},
	note = {Publisher: American Chemical Society},
}

@article{zhao_ultrahigh-mobility_2024,
	title = {Ultrahigh-mobility semiconducting epitaxial graphene on silicon carbide},
	volume = {625},
	rights = {2024 The Author(s), under exclusive licence to Springer Nature Limited},
	issn = {1476-4687},
	url = {https://www.nature.com/articles/s41586-023-06811-0},
	doi = {10.1038/s41586-023-06811-0},
	pages = {60--65},
	number = {7993},
	journaltitle = {Nature},
	author = {Zhao, Jian and Ji, Peixuan and Li, Yaqi and Li, Rui and Zhang, Kaimin and Tian, Hao and Yu, Kaicheng and Bian, Boyue and Hao, Luzhen and Xiao, Xue and Griffin, Will and Dudeck, Noel and Moro, Ramiro and Ma, Lei and de Heer, Walt A.},
	urldate = {2025-10-31},
	date = {2024-01},
	langid = {english},
	note = {Publisher: Nature Publishing Group},
}

@article{n_nair_band_2017,
	title = {Band Gap Opening Induced by the Structural Periodicity in Epitaxial Graphene Buffer Layer},
	volume = {17},
	issn = {1530-6984},
	url = {https://doi.org/10.1021/acs.nanolett.7b00509},
	doi = {10.1021/acs.nanolett.7b00509},
	pages = {2681--2689},
	number = {4},
	journaltitle = {Nano Letters},
	shortjournal = {Nano Lett.},
	author = {N. Nair, Maya and Palacio, Irene and Celis, Arlensiú and Zobelli, Alberto and Gloter, Alexandre and Kubsky, Stefan and Turmaud, Jean-Philippe and Conrad, Matthew and Berger, Claire and de Heer, Walter and Conrad, Edward H. and Taleb-Ibrahimi, Amina and Tejeda, Antonio},
	urldate = {2025-10-31},
	date = {2017-04-12},
	note = {Publisher: American Chemical Society},
}

@article{xu_interfacial_2018,
	title = {Interfacial engineering in graphene bandgap},
	volume = {47},
	issn = {0306-0012, 1460-4744},
	url = {https://xlink.rsc.org/?DOI=C7CS00836H},
	doi = {10.1039/C7CS00836H},
	pages = {3059--3099},
	number = {9},
	journaltitle = {Chemical Society Reviews},
	shortjournal = {Chem. Soc. Rev.},
	author = {Xu, Xiaozhi and Liu, Chang and Sun, Zhanghao and Cao, Ting and Zhang, Zhihong and Wang, Enge and Liu, Zhongfan and Liu, Kaihui},
	urldate = {2025-10-31},
	date = {2018},
	langid = {english},
}

@article{avetisyan_stacking_2010,
	title = {Stacking order dependent electric field tuning of the band gap in graphene multilayers},
	volume = {81},
	url = {https://link.aps.org/doi/10.1103/PhysRevB.81.115432},
	doi = {10.1103/PhysRevB.81.115432},
	pages = {115432},
	number = {11},
	journaltitle = {Physical Review B},
	shortjournal = {Phys. Rev. B},
	author = {Avetisyan, A. A. and Partoens, B. and Peeters, F. M.},
	urldate = {2025-10-31},
	date = {2010-03-18},
	note = {Publisher: American Physical Society},
}

@article{wong_strain_2012,
	title = {Strain Effect on the Electronic Properties of Single Layer and Bilayer Graphene},
	volume = {116},
	issn = {1932-7447},
	url = {https://doi.org/10.1021/jp300840k},
	doi = {10.1021/jp300840k},
	pages = {8271--8277},
	number = {14},
	journaltitle = {The Journal of Physical Chemistry C},
	shortjournal = {J. Phys. Chem. C},
	author = {Wong, Jen-Hsien and Wu, Bi-Ru and Lin, Ming-Fa},
	urldate = {2025-10-31},
	date = {2012-04-12},
	note = {Publisher: American Chemical Society},
}

@article{xia_graphene_2010,
	title = {Graphene Field-Effect Transistors with High On/Off Current Ratio and Large Transport Band Gap at Room Temperature},
	volume = {10},
	issn = {1530-6984},
	url = {https://doi.org/10.1021/nl9039636},
	doi = {10.1021/nl9039636},
	pages = {715--718},
	number = {2},
	journaltitle = {Nano Letters},
	shortjournal = {Nano Lett.},
	author = {Xia, Fengnian and Farmer, Damon B. and Lin, Yu-ming and Avouris, Phaedon},
	urldate = {2025-10-31},
	date = {2010-02-10},
	note = {Publisher: American Chemical Society},
}

@article{mak_observation_2009,
	title = {Observation of an Electric-Field-Induced Band Gap in Bilayer Graphene by Infrared Spectroscopy},
	volume = {102},
	url = {https://link.aps.org/doi/10.1103/PhysRevLett.102.256405},
	doi = {10.1103/PhysRevLett.102.256405},
	pages = {256405},
	number = {25},
	journaltitle = {Physical Review Letters},
	shortjournal = {Phys. Rev. Lett.},
	author = {Mak, Kin Fai and Lui, Chun Hung and Shan, Jie and Heinz, Tony F.},
	urldate = {2025-10-31},
	date = {2009-06-26},
	note = {Publisher: American Physical Society},
}

@article{gui_reply_2009,
	title = {Reply to ``Comment on `Band structure engineering of graphene by strain: First-principles calculations' ''},
	volume = {80},
	url = {https://link.aps.org/doi/10.1103/PhysRevB.80.167402},
	doi = {10.1103/PhysRevB.80.167402},
	shorttitle = {Reply to ``Comment on `Band structure engineering of graphene by strain},
	pages = {167402},
	number = {16},
	journaltitle = {Physical Review B},
	shortjournal = {Phys. Rev. B},
	author = {Gui, Gui and Li, Jin and Zhong, Jianxin},
	urldate = {2025-10-31},
	date = {2009-10-19},
	note = {Publisher: American Physical Society},
}

@article{ni_uniaxial_2009,
	title = {Uniaxial Strain on Graphene: Raman Spectroscopy Study and Band-Gap Opening},
	volume = {3},
	issn = {1936-0851},
	url = {https://doi.org/10.1021/nn8008323},
	doi = {10.1021/nn8008323},
	shorttitle = {Uniaxial Strain on Graphene},
	pages = {483--483},
	number = {2},
	journaltitle = {{ACS} Nano},
	shortjournal = {{ACS} Nano},
	author = {Ni, Zhen Hua and Yu, Ting and Lu, Yun Hao and Wang, Ying Ying and Feng, Yuan Ping and Shen, Ze Xiang},
	urldate = {2025-10-31},
	date = {2009-02-24},
	note = {Publisher: American Chemical Society},
}

@article{pereira_tight-binding_2009,
	title = {Tight-binding approach to uniaxial strain in graphene},
	volume = {80},
	url = {https://link.aps.org/doi/10.1103/PhysRevB.80.045401},
	doi = {10.1103/PhysRevB.80.045401},
	pages = {045401},
	number = {4},
	journaltitle = {Physical Review B},
	shortjournal = {Phys. Rev. B},
	author = {Pereira, Vitor M. and Castro Neto, A. H. and Peres, N. M. R.},
	urldate = {2025-10-31},
	date = {2009-07-01},
	note = {Publisher: American Physical Society},
}

@misc{vaswani_attention_2023,
	title = {Attention Is All You Need},
	url = {http://arxiv.org/abs/1706.03762},
	doi = {10.48550/arXiv.1706.03762},
	number = {{arXiv}:1706.03762},
	publisher = {{arXiv}},
	author = {Vaswani, Ashish and Shazeer, Noam and Parmar, Niki and Uszkoreit, Jakob and Jones, Llion and Gomez, Aidan N. and Kaiser, Lukasz and Polosukhin, Illia},
	urldate = {2025-10-16},
	date = {2023-08-02},
	eprinttype = {arxiv},
	eprint = {1706.03762 [cs]},
}

@misc{kingma_adam_2017,
	title = {Adam: A Method for Stochastic Optimization},
	url = {http://arxiv.org/abs/1412.6980},
	doi = {10.48550/arXiv.1412.6980},
	shorttitle = {Adam},
	number = {{arXiv}:1412.6980},
	publisher = {{arXiv}},
	author = {Kingma, Diederik P. and Ba, Jimmy},
	urldate = {2025-10-16},
	date = {2017-01-30},
	eprinttype = {arxiv},
	eprint = {1412.6980 [cs]},
}

@article{sabattini_towards_2025,
	title = {Towards {AI}-driven autonomous growth of 2D materials based on a graphene case study},
	volume = {8},
	rights = {2025 The Author(s)},
	issn = {2399-3650},
	url = {https://www.nature.com/articles/s42005-025-02086-1},
	doi = {10.1038/s42005-025-02086-1},
	pages = {180},
	number = {1},
	journaltitle = {Communications Physics},
	shortjournal = {Commun Phys},
	author = {Sabattini, Leonardo and Coriolano, Annalisa and Casert, Corneel and Forti, Stiven and Barnard, Edward S. and Beltram, Fabio and Pontil, Massimiliano and Whitelam, Stephen and Coletti, Camilla and Rossi, Antonio},
	urldate = {2025-06-18},
	date = {2025-04-25},
	langid = {english},
	note = {Publisher: Nature Publishing Group},
}

@article{whitaker_simple_2018,
	title = {A simple algorithm for despiking Raman spectra},
	volume = {179},
	issn = {0169-7439},
	url = {https://www.sciencedirect.com/science/article/pii/S0169743918301758},
	doi = {10.1016/j.chemolab.2018.06.009},
	pages = {82--84},
	journaltitle = {Chemometrics and Intelligent Laboratory Systems},
	shortjournal = {Chemometrics and Intelligent Laboratory Systems},
	author = {Whitaker, Darren A. and Hayes, Kevin},
	urldate = {2025-08-20},
	date = {2018-08-15},
}

@article{nakashima_raman_1997,
	title = {Raman Investigation of {SiC} Polytypes},
	volume = {162},
	rights = {Copyright © 1997 {WILEY}-{VCH} Verlag Berlin {GmbH}, Fed. Rep. of Germany},
	issn = {1521-396X},
	url = {https://onlinelibrary.wiley.com/doi/abs/10.1002/1521-396X%28199707%29162%3A1%3C39%3A%3AAID-PSSA39%3E3.0.CO%3B2-L},
	doi = {10.1002/1521-396X(199707)162:1<39::AID-PSSA39>3.0.CO;2-L},
	pages = {39--64},
	number = {1},
	journaltitle = {physica status solidi (a)},
	author = {Nakashima, S. and Harima, H.},
	urldate = {2025-07-04},
	date = {1997},
	langid = {english},
	note = {\_eprint: https://onlinelibrary.wiley.com/doi/pdf/10.1002/1521-396X\%28199707\%29162\%3A1\%3C39\%3A\%3AAID-{PSSA}39\%3E3.0.{CO}\%3B2-L},
}

@article{radtke_vibrational_2025,
	title = {Vibrational properties of epitaxial graphene buffer layer on silicon carbide},
	volume = {111},
	url = {https://link.aps.org/doi/10.1103/xk69-4n46},
	doi = {10.1103/xk69-4n46},
	pages = {L220303},
	number = {22},
	journaltitle = {Physical Review B},
	shortjournal = {Phys. Rev. B},
	author = {Radtke, Guillaume and Lazzeri, Michele},
	urldate = {2025-06-24},
	date = {2025-06-16},
	note = {Publisher: American Physical Society},
}

@article{fromm_contribution_2013,
	title = {Contribution of the buffer layer to the Raman spectrum of epitaxial graphene on {SiC}(0001)},
	volume = {15},
	issn = {1367-2630},
	url = {https://dx.doi.org/10.1088/1367-2630/15/4/043031},
	doi = {10.1088/1367-2630/15/4/043031},
	pages = {043031},
	number = {4},
	journaltitle = {New Journal of Physics},
	shortjournal = {New J. Phys.},
	author = {Fromm, F and Oliveira Jr, M H and Molina-Sánchez, A and Hundhausen, M and Lopes, J M J and Riechert, H and Wirtz, L and Seyller, T},
	urldate = {2025-06-18},
	date = {2013-04},
	langid = {english},
	note = {Publisher: {IOP} Publishing},
}

@article{Cavallucci2018,
  title = {Intrinsic structural and electronic properties of the Buffer Layer on Silicon Carbide unraveled by Density Functional Theory},
  volume = {8},
  ISSN = {2045-2322},
  url = {http://dx.doi.org/10.1038/s41598-018-31490-7},
  DOI = {10.1038/s41598-018-31490-7},
  number = {1},
  journal = {Scientific Reports},
  publisher = {Springer Science and Business Media LLC},
  author = {Cavallucci,  Tommaso and Tozzini,  Valentina},
  year = {2018},
  month = aug 
}

@article{turner2023introduction,
  title={An introduction to transformers},
  author={Turner, Richard E},
  journal={arXiv preprint arXiv:2304.10557},
  year={2023}
}

@article{Cavallucci2018Carbon,
  author  = {Cavallucci, Tommaso and Murata, Yuya and Takamura, Makoto and Hibino, Hiroki and Heun, Stefan and Tozzini, Valentina},
  title   = {Unraveling localized states in quasi free standing monolayer graphene by means of Density Functional Theory},
  journal = {Carbon},
  year    = {2018},
  volume  = {130},
  pages   = {466-474},
  doi     = {10.1016/j.carbon.2018.01.027}
}

@article{Cavallucci2016,
  title = {Multistable Rippling of Graphene on SiC: A Density Functional Theory Study},
  volume = {120},
  ISSN = {1932-7455},
  url = {http://dx.doi.org/10.1021/acs.jpcc.6b01356},
  DOI = {10.1021/acs.jpcc.6b01356},
  number = {14},
  journal = {The Journal of Physical Chemistry C},
  publisher = {American Chemical Society (ACS)},
  author = {Cavallucci,  Tommaso and Tozzini,  Valentina},
  year = {2016},
  month = apr,
  pages = {7670–7677}
}

@article{Rappe1990PRB,
  author  = {Rappe, Andrew M. and Rabe, Karin M. and Kaxiras, Efthimios and Joannopoulos, John D.},
  title   = {Optimized pseudopotentials},
  journal = {Physical Review B},
  year    = {1990},
  volume  = {41},
  pages   = {1227-1230},
  doi     = {10.1103/PhysRevB.41.1227}
}

@article{Perdew1996PRL,
  author  = {Perdew, John P. and Burke, Kieron and Ernzerhof, Matthias},
  title   = {Generalized Gradient Approximation Made Simple},
  journal = {Physical Review Letters},
  year    = {1996},
  volume  = {77},
  pages   = {3865-3868},
  doi     = {10.1103/PhysRevLett.77.3865}
}

@article{Grimme2006JCC,
  author  = {Grimme, Stefan},
  title   = {Semiempirical GGA-type density functional constructed with a long-range dispersion correction},
  journal = {Journal of Computational Chemistry},
  year    = {2006},
  volume  = {27},
  pages   = {1787-1799},
  doi     = {10.1002/jcc.20495}
}

@article{Billeter2003CMS,
  author  = {Billeter, Salomon R. and Curioni, Alessandro and Andreoni, Wanda},
  title   = {Efficient linear scaling geometry optimization and transition-state search for direct wavefunction optimization schemes in density functional theory using a plane-wave basis},
  journal = {Computational Materials Science},
  year    = {2003},
  volume  = {27},
  pages   = {437-445},
  doi     = {10.1016/S0927-0256(03)00043-0}
}

@article{Giannozzi2009JPCM,
  author  = {Giannozzi, Paolo and Baroni, Stefano and Bonini, Nicola and Calandra, Matteo and Car, Roberto and Cavazzoni, Carlo and Ceresoli, Davide and Chiarotti, Gian L. and Cococcioni, Matteo and {et al.}},
  title   = {QUANTUM ESPRESSO: a modular and open-source software project for quantum simulations of materials},
  journal = {Journal of Physics: Condensed Matter},
  year    = {2009},
  volume  = {21},
  pages   = {395502},
  doi     = {10.1088/0953-8984/21/39/395502}
}

@article{alamode,
  author = {Tadano, T. and Gohda, Y. and Tsuneyuki, S.},
  title = {Anharmonic force constants extracted from first-principles molecular dynamics: Applications to heat transfer simulations},
  journal = {Journal of Physics: Condensed Matter},
  volume = {26},
  number = {22},
  pages = {225402},
  year = {2014},
  doi = {10.1088/0953-8984/26/22/225402}
}

@article{Radtke2025,
  title = {Vibrational properties of epitaxial graphene buffer layer on silicon carbide},
  volume = {111},
  ISSN = {2469-9969},
  url = {http://dx.doi.org/10.1103/xk69-4n46},
  DOI = {10.1103/xk69-4n46},
  number = {22},
  journal = {Physical Review B},
  publisher = {American Physical Society (APS)},
  author = {Radtke,  Guillaume and Lazzeri,  Michele},
  year = {2025},
  month = jun 
}

@article{Mounet2005,
  title = {First-principles determination of the structural,  vibrational and thermodynamic properties of diamond,  graphite,  and derivatives},
  volume = {71},
  ISSN = {1550-235X},
  url = {http://dx.doi.org/10.1103/PhysRevB.71.205214},
  DOI = {10.1103/physrevb.71.205214},
  number = {20},
  journal = {Physical Review B},
  publisher = {American Physical Society (APS)},
  author = {Mounet,  Nicolas and Marzari,  Nicola},
  year = {2005},
  month = may 
}

@article{Nakano2021,
  title = {Atomic forces by quantum Monte Carlo: Application to phonon dispersion calculations},
  volume = {103},
  ISSN = {2469-9969},
  url = {http://dx.doi.org/10.1103/PhysRevB.103.L121110},
  DOI = {10.1103/physrevb.103.l121110},
  number = {12},
  journal = {Physical Review B},
  publisher = {American Physical Society (APS)},
  author = {Nakano,  Kousuke and Morresi,  Tommaso and Casula,  Michele and Maezono,  Ryo and Sorella,  Sandro},
  year = {2021},
  month = mar 
}

@article{Tuinstra1970,
  author  = {Tuinstra, F. and Koenig, J. L.},
  title   = {Raman Spectrum of Graphite},
  journal = {The Journal of Chemical Physics},
  volume  = {53},
  pages   = {1126-1130},
  year    = {1970},
  doi     = {10.1063/1.1674108}
}

@article{Venezuela2011,
  author  = {Venezuela, P. and Lazzeri, M. and Mauri, F.},
  title   = {Theory of double-resonant Raman spectra in graphene: Intensity and line shape of defect-induced and two-phonon bands},
  journal = {Physical Review B},
  volume  = {84},
  pages   = {035433},
  year    = {2011},
  doi     = {10.1103/PhysRevB.84.035433}
}

@article{Maultzsch2004,
  author  = {Maultzsch, J. and Reich, S. and Thomsen, C.},
  title   = {Double-resonant Raman scattering in graphite: Interference effects, selection rules, and phonon dispersion},
  journal = {Physical Review B},
  volume  = {70},
  pages   = {155403},
  year    = {2004},
  doi     = {10.1103/PhysRevB.70.155403}
}

@article{Ferrari2013,
  author  = {Ferrari, Andrea C. and Basko, Denis M.},
  title   = {Raman spectroscopy as a versatile tool for studying the properties of graphene},
  journal = {Nature Nanotechnology},
  volume  = {8},
  pages   = {235-246},
  year    = {2013},
  doi     = {10.1038/nnano.2013.46}
}

@article{Ferrari2000,
  title = {Interpretation of Raman spectra of disordered and amorphous carbon},
  volume = {61},
  ISSN = {1095-3795},
  url = {http://dx.doi.org/10.1103/PhysRevB.61.14095},
  DOI = {10.1103/physrevb.61.14095},
  number = {20},
  journal = {Physical Review B},
  publisher = {American Physical Society (APS)},
  author = {Ferrari,  A. C. and Robertson,  J.},
  year = {2000},
  month = may,
  pages = {14095–14107}
}
\end{refsection}


\end{document}